\documentclass[a4paper,twoside,openany,11pt]{memoir} 
\usepackage{minibox}
\usepackage{longtable}
\def\fullheadfoot{0}

\usepackage[english]{babel}
\usepackage{microtype,xspace}
\usepackage[utf8]{inputenc}	

\usepackage{amsfonts,amsmath,amssymb,bm,mathrsfs,mathtools,dsfont} 	
\allowdisplaybreaks 	
\usepackage[table,dvipsnames]{xcolor}
\usepackage{hyperref}
\usepackage{slashed}
\usepackage{multicol}

\usepackage{siunitx}
\usepackage{geometry}
\usepackage[sort&compress,numbers,merge]{natbib}
\usepackage{chapterbib}
\addto\captionsenglish{\renewcommand*{\bibname}{References}}
\makeatletter
\renewcommand{\@memb@bchap}{ 
\bibmark \prebibhook
}
\makeatother

\usepackage{graphicx}
\graphicspath{{./Figures/}}
\usepackage{caption,subcaption}
\usepackage{multirow}
\renewcommand{\arraystretch}{1.2}
\usepackage{tabularx}
\newcolumntype{Y}{>{\centering\arraybackslash}X}

\usepackage[shortlabels]{enumitem}
\setlist{itemsep=.1em,topsep=.5em}
\SetEnumerateShortLabel{i}{\textit{\roman*}}

\definecolor{red}{rgb}{0.6,.0706,.1373}
\definecolor{blue}{rgb}{0,0.396,0.741}
\definecolor{green}{rgb}{0.25,0.6,0.2}
\definecolor{teal}{rgb}{0.11,0.6,0.6}
\definecolor{orange}{rgb}{.8, .4806, 0.173}
\definecolor{yellow}{rgb}{.8, .7, 0.05}
\colorlet{blueref}{blue!80!black}
\colorlet{bluelink}{blue!90!black}
\hypersetup{
	colorlinks, 
	bookmarksopen, 
	bookmarksnumbered,
	citecolor=blueref, 		
	linkcolor=bluelink,	
	urlcolor=bluelink,			
	pdftitle={Evanescent Operators in One-Loop Matching Computations},    
}

\addto\captionsenglish{}

\renewcommand{\contentsname}{Contents}
\renewcommand{\printtoctitle}[1]{}

\settocdepth{subsection}
\newcommand{\toc}{ {
	\hypersetup{linkcolor = black} 
	\vspace*{-.06\textheight}	
	\tableofcontents*
	\thispagestyle{standardstyle} 
} }

\makeatletter
\newcommand*\ifthispageodd{%
  \checkoddpage
  \ifoddpage
    \expandafter\@firstoftwo
  \else
    \expandafter\@secondoftwo
  \fi
}
\makeatother

\usepackage[noindentafter]{titlesec} 
\counterwithout{section}{chapter} 
\counterwithout{figure}{chapter} 
\counterwithout{table}{chapter} 
\numberwithin{equation}{section} 
\setsecnumdepth{subsubsection}

\DeclareMathVersion{sans}
\SetSymbolFont{operators}{sans}{OT1}{cmbr}{m}{n}
\SetSymbolFont{letters}{sans}{OML}{cmbrm}{m}{it}
\SetSymbolFont{symbols}{sans}{OMS}{cmbrs}{m}{n}
\SetMathAlphabet{\mathit}{sans}{OT1}{cmbr}{m}{sl}
\SetMathAlphabet{\mathbf}{sans}{OT1}{cmbr}{bx}{n}
\SetMathAlphabet{\mathtt}{sans}{OT1}{cmtl}{m}{n}
\SetSymbolFont{largesymbols}{sans}{OMX}{iwona}{m}{n}

\DeclareMathVersion{boldsans}
\SetSymbolFont{operators}{boldsans}{OT1}{cmbr}{b}{n}
\SetSymbolFont{letters}{boldsans}{OML}{cmbrm}{b}{it}
\SetSymbolFont{symbols}{boldsans}{OMS}{cmbrs}{b}{n}
\SetMathAlphabet{\mathit}{boldsans}{OT1}{cmbr}{b}{sl}
\SetMathAlphabet{\mathbf}{boldsans}{OT1}{cmbr}{bx}{n}
\SetMathAlphabet{\mathtt}{boldsans}{OT1}{cmtl}{b}{n}
\SetSymbolFont{largesymbols}{boldsans}{OMX}{iwona}{bx}{n}

\titleformat{\section}{\centering \Large \bfseries \sffamily \mathversion{boldsans} \color{blue!80!black} }{\thesection}{15pt}{}{}
\titlespacing{\section}{0pt}{15pt}{5pt}
\titleformat{\subsection}{\large \sffamily \mathversion{sans} \color{blue!70!black} }{\thesubsection}{10pt}{}{}
\titlespacing{\subsection}{0pt}{10pt}{5pt}
\titleformat{\subsubsection}{\normalsize \sffamily \itshape \mathversion{sans} \color{blue!70!black} }{\thesubsubsection}{10pt}{}{}
\titlespacing{\subsubsection}{0pt}{10pt}{0pt}

\ifnum\fullheadfoot=1
	\makepagestyle{standardstyle}
	\makeoddhead{standardstyle}{\sffamily \mathversion{subsectionmath} \rightmark}{}{\sffamily \bfseries{\thepage}}
	\makeevenhead{standardstyle}{\sffamily \bfseries{\thepage}}{}{\sffamily \mathversion{subsectionmath} \leftmark}
	\makeheadrule{standardstyle}{\textwidth}{\normalrulethickness}
	\makepsmarks{standardstyle}{
		\nouppercaseheads
		\createmark{section}{both}{shownumber}{\ }{\hspace{1.2em}  }
		\createmark{subsection}{right}{shownumber}{}{\hspace{1.2em} }
	}
	
	\copypagestyle{appstyle}{standardstyle}
	\makeevenhead{appstyle}{\sffamily \bfseries{\thepage}}{}{ \sffamily \mathversion{subsectionmath} \leftmark}
	\makepsmarks{appstyle}{
		\nouppercaseheads
		\createmark{section}{both}{nonumber}{\ }{\ \ }
		\createmark{subsection}{right}{shownumber}{}{\ \ }
	}
\else 
	\makepagestyle{standardstyle}
	\makeoddfoot{standardstyle}{}{\sffamily -- {\bfseries\thepage} --}{} 
	\makeevenfoot{standardstyle}{}{\sffamily -- {\bfseries\thepage} --}{}
	\copypagestyle{appstyle}{standardstyle}
\fi

\createplainmark{toc}{both}{\contentsname}
\createplainmark{bib}{both}{\bibname}

\makeatletter
\let\MyIntOrig\int
\def\MyIntSpace{\hspace{-.35em}} 
\def\int{\MyInt}
\def\MyInt{\MyIntOrig\MyIntSkipMaybe}
\def\MyIntSkipMaybe{
	\@ifnextchar_{\MyIntSkipScript}{%
		\@ifnextchar^{\MyIntSkipScript}{%
			\@ifnextchar\limits{\MyIntSkipTok}{%
				\@ifnextchar\nolimits{\MyIntSkipTok}{%
					\MyIntSpace}}}}%
}
\def\MyIntSkipScript#1#2{#1{#2}\MyIntSkipMaybe}
\def\MyIntSkipTok#1{#1\MyIntSkipMaybe}

\newcommand{\pushright}[1]{\ifmeasuring@#1\else\omit\hfill$\displaystyle#1$\fi\ignorespaces}
\makeatother

\usepackage{bbm}
\RequirePackage{fix-cm}
\usepackage{fontawesome}
\usepackage{booktabs}
\usepackage{multirow}
\usepackage{soul}
\usepackage{bbold}

\newcommand{\tr}{\mathop{\mathrm{tr}} }
\newcommand{\eminus}{\vcenter{\hbox{\scalebox{0.6}[1]{$ - $}}}}	
\newcommand{\ord}[1]{\mathcal{O}( #1 )}
\newcommand{\commutator}[2]{\big[#1, \, #2 \big]}

\newcommand{\dd}{\mathop{}\!\mathrm{d}}
\newcommand{\ud}[2]{\phantom{}^{#1}\phantom{}_{#2}}

\newcommand{\rep}[1]{\mathbf{#1}}

\newcommand{\sscript}[1]{{\scriptscriptstyle \mathrm{#1}}}

\renewcommand{\L}{\mathcal{L}}

\newcommand{\U}{\mathrm{U}}

\newcommand{\UV}{\sscript{UV}}
\newcommand{\EFT}{\sscript{EFT}}

\newcommand{\str}{\mathop{\mathrm{STr}}}

\newcommand{\bef}{$ \beta $-function\xspace}
\newcommand{\befs}{$ \beta $-functions\xspace}
\newcommand{\msbar}{$ \overline{\text{\small MS}} $\xspace}
\newcommand{\ms}{$ \text{\small MS} $\xspace}

\definecolor{verde}{cmyk}{0.92,0,0.59,0.25}

\usepackage[normalem]{ulem}

\usepackage[utf8]{inputenc}

\begin{document}

\thispagestyle{empty}
\renewcommand*{\thefootnote}{\fnsymbol{footnote}}
\begin{center} 
\begin{minipage}{15.5cm}
\vspace{-0.7cm}
\begin{flushright}
{\footnotesize \itshape
MITP-22-091\\
TUM-HEP-1428/22\\[-3pt]
ZU-TH-48/22
}
\end{flushright}

\end{minipage}
\end{center}

\begin{center}
	{\sffamily \bfseries \fontsize{16.}{20}\selectfont \mathversion{boldsans}
	Evanescent Operators in One-Loop Matching Computations\\[-.5em]
	\textcolor{blue!80!black}{\rule{\textwidth}{2pt}}\\
	\vspace{.05\textheight}}
	{\sffamily \mathversion{sans} \Large Javier Fuentes-Mart\'{\i}n,$^{1,2}$\footnote{javier.fuentes@ugr.es} 
	Matthias König,$^{3}$\footnote{matthias.koenig@tum.de}
	Julie Pagès,$^{4}$\footnote{julie.pages@physics.ucsd.edu}\\[5pt]
	Anders Eller Thomsen,$^{5}$\footnote{thomsen@itp.unibe.ch}
	and Felix Wilsch$^{6}$\footnote{felix.wilsch@physik.uzh.ch}}
	\\[1.25em]
	{ \small \sffamily \mathversion{sans} 
	$^{1}\,$  Departamento de F\'isica Te\'orica y del Cosmos, Universidad de Granada,\\
    Campus de Fuentenueva, E–18071 Granada, Spain,\\[5pt]
    $^{2}\,$  PRISMA Cluster of Excellence \& Mainz Institute for Theoretical Physics,\\
    Johannes Gutenberg University, D-55099 Mainz, Germany,\\[5pt]
	$^{3}\,$ Physik Department T31, Technische Universit\"at M\"unchen,\\
	James-Franck-Str. 1, D-85748 Garching, Germany\\[5pt]
	$^{4}\,$ Department of Physics, University of California at San Diego, \\9500 Gilman Drive,
    La Jolla, CA 92093-0319, USA\\[5pt]
	$^{5}\,$ Albert Einstein Center for Fundamental Physics, Institute for Theoretical Physics,\\ University of Bern, CH-3012 Bern, Switzerland\\[5pt]
	$^{6}\,$ Physik-Institut, Universit\"at Z\"urich, CH-8057 Z\"urich, Switzerland
	}
	\\[.005\textheight]{\itshape \sffamily \today}
	\\[.03\textheight]
\end{center}
\setcounter{footnote}{0}
\renewcommand*{\thefootnote}{\arabic{footnote}}%
\suppressfloats	

\begin{abstract}\vspace{+.01\textheight}
Effective Field Theory calculations used in countless phenomenological analyses employ dimensional regularization, and at intermediate stages of computations, the operator bases extend beyond the four-dimensional ones. The extra pieces---the evanescent operators---can ultimately be removed with a suitable renormalization scheme, resulting in a finite shift of the physical operators. Modern Effective Field Theory matching techniques relying on the method of expansion by regions have to be extended to account for this. After illustrating the importance of these shifts in two specific examples, we compute the finite shifts required to remove all evanescent operators appearing in the one-loop matching of generic ultraviolet theories to the Standard Model Effective Field Theory and elucidate the formalism for generic Effective Field Theory calculations.
\end{abstract}

\newpage
\section*{Table of Contents}
\toc
\newpage

\section{Introduction}
\label{sec:intro}

Low-energy phenomenology has long been crucial to the search for new particles in and beyond the Standard Model (SM). It is well known that the presence of heavy particles can be inferred before they become directly accessible through their indirect effect on low-energy processes. With the apparent absence of new physics~(NP) around the electroweak~(EW) scale, and the appearance of low-energy hints of deviations from the SM predictions~\cite{deSimone:2020kwi,Muong-2:2021ojo}, indirect searches are progressively taking a more prominent role in the Beyond the Standard Model~(BSM) program. Typically, when exploring the low-energy physics of BSM models, one matches the models onto the SM Effective Field Theory (SMEFT)~\cite{Buchmuller:1985jz,Grzadkowski:2010es}, the appropriate Effective Field Theory (EFT) for weakly-coupled models in the absence of light new states. The SMEFT is then run down to lower energies using its known Renormalization Group~(RG) equations~\cite{Jenkins:2013zja,Jenkins:2013wua,Alonso:2013hga,Alonso:2014zka} until the EW scale. To compute the effects on, e.g., flavor observables, the EFT is further matched to the Low-Energy Field Theory (LEFT)~\cite{Jenkins:2017jig,Dekens:2019ept,Aebischer:2015fzz} and evolved down to the appropriate energies with the LEFT RG equations~\cite{Jenkins:2017dyc}. 

Many tools have already been developed to automate these computations within the SMEFT-to-LEFT framework~\cite{Celis:2017hod,Aebischer:2018bkb,Fuentes-Martin:2020zaz,Criado:2017khh,Brivio:2019irc,Gripaios:2018zrz,Criado:2019ugp,Dedes:2019uzs,Hartland:2019bjb,Aebischer:2018iyb,EOSAuthors:2021xpv,Straub:2018kue,Brivio:2017btx,Uhlrich:2020ltd,Bakshi:2018ics,DiNoi:2022ejg}, making these computations accessible to the community at large. Moreover, much work has been done to develop suitable methods for one-loop matching of arbitrary BSM models to their EFTs with the functional approach~\cite{Gaillard:1985uh,Chan:1986jq,Cheyette:1987qz,Chan:1985ny,Fraser:1984zb,Aitchison:1984ys,Aitchison:1985pp,Aitchison:1985hu,Cheyette:1985ue,Chan:1986jq,Dittmaier:1995cr,Dittmaier:1995ee,Henning:2014wua,Drozd:2015rsp,delAguila:2016zcb,Boggia:2016asg,Henning:2016lyp,Ellis:2016enq,Fuentes-Martin:2016uol,Zhang:2016pja,Ellis:2017jns,Summ:2018oko,Cohen:2019btp,Cohen:2020fcu}, resulting in several partial one-loop EFT matching results~\cite{Kramer:2019fwz,Angelescu:2020yzf,Ellis:2020ivx,Dedes:2021abc} as well as tools to automate the tedious and demanding task of evaluating functional supertraces at the center of the functional matching procedure~\cite{Cohen:2020qvb,Fuentes-Martin:2020udw}. A new generation of tools is now aiming at solving the more generic problem of completely automating one-loop matching and running using diagrammatic~\cite{Carmona:2021xtq} and functional methods~\cite{Matchete}.

Despite the recent progress, the incorporation of one-loop corrections in the EFT matching, crucial in many phenomenological analyses, still presents a great challenge. These calculations are typically carried out using dimensional regularization (DR) due to its convenience and simplicity. Nevertheless, loop computations using DR need to account for several subtitles such as the presence of \emph{evanescent operators}.\footnote{The presence of evanescent operators is not related to the issue of extending four-dimensional objects such as $ \gamma_5 $ and $ \varepsilon_{\mu\nu\rho\sigma} $ to $d$ dimensions and presents a separate, independent challenge~\cite{Dugan:1990df}.} In $d= 4 - 2 \epsilon$ dimensions, the Lorentz algebra is infinite-dimensional and the four-dimensional operator bases for the EFT do not span the operator space in $d$ dimensions. Conversely, $d$-dimensional operators cannot be expressed in terms of the four-dimensional basis operators using standard relations such as Fierz identities, since these are valid only in strictly four dimensions. In short, reducing the $d$-dimensional EFT to a suitable four-dimensional basis leaves remnants---evanescent operators---which cannot be neglected although they vanish in the four-dimensional limit. The evanescent operators can formally be considered of $\mathcal{O}(\epsilon)$ and do not contribute to tree-level amplitudes. However, when inserted in loop amplitudes, they yield additional finite contributions, which need to be accounted for beyond leading-order computations. It turns out that it is not necessary to keep track of all the evanescent operators throughout EFT calculations. In physical amplitudes, the evanescent operators give finite contributions only when picking out the poles of the loop diagrams. Their impact is therefore local and can be compensated by suitable counterterms. This lets us define renormalization schemes completely free of all contributions from evanescent operators, which are effectively eliminated from the EFT. 

Evanescent operators were originally studied in the context of next-to-leading order (NLO) calculations of anomalous dimensions in the Weak Effective Hamiltonian~\cite{Buras:1989xd,Dugan:1990df,Herrlich:1994kh} and have since been included in many other NLO computations of SM processes. Recently, they have also been encountered in one-loop basis changes in the LEFT~\cite{Aebischer:2022tvz,Aebischer:2022aze,Aebischer:2022rxf}. However, contributions from evanescent operators are easily missed when performing one-loop matching computations, in particular when matching NP models onto the SMEFT. Modern matching techniques rely on the method of regions~\cite{Beneke:1997zp,Jantzen:2011nz} to greatly simplify the computations. These eliminate the need for computing EFT contributions by considering only ultraviolet~(UV) loops in the hard region~\cite{Fuentes-Martin:2016uol,Zhang:2016pja}, where the loop momentum is of the order of the heavy scales. Nevertheless, to determine the EFT action in a scheme where the evanescent contributions can be disregarded, the additional computation of finite counterterms is required. The finite counterterms of such non-minimal schemes cancel certain loops in the EFT by construction. Thus, a loop computation \emph{in the EFT} is necessary to determine them. Once the counterterms have been calculated, the renormalized couplings can be recovered as the difference between the bare couplings and the counterterms. A simple procedure to recover the physical EFT action $S_\EFT^S$, with no contributions from evanescent operators, is given by the following prescription:
\begin{enumerate}[i)]
	\item Compute the bare EFT action, $S_\EFT$, using the method of regions. 
	\item Decompose the operator basis into a physical and an evanescent part. This decomposition is not unique, thus requiring a prescription for the projection to the physical basis.
	\item Remove the evanescent part from the tree-level EFT action, $ S_\EFT^{(0)} $, to construct its physical counterpart, $ S_\EFT^{S(0)} $.
	\item The one-loop physical EFT action, $ S_\EFT^{S(1)}$, is obtained by shifting the one-loop bare action $S_\EFT^{(1)}$ with the difference between one-particle irreducible~(1PI) loops generated by the tree-level actions $ S_\EFT^{(0)} $  and $ S_\EFT^{S(0)} $, which compensates for the finite contributions from the evanescent operators. 
\end{enumerate}
In the last step, it is sufficient to extract the $ 1/\epsilon $ UV poles from the loop integrals, since the insertion of evanescent operators contributes with a factor $ \epsilon $. This renders the computation entirely elementary. 

In this paper we work our way from the specific to the abstract in the treatment of evanescent operators in generic EFTs. In Section~\ref{sec:pheno}, we examine a concrete example of an evanescent contribution to the dipole operators showing up in the matching of the Two-Higgs-Doublet Model (2HDM) to the SMEFT Warsaw basis~\cite{Grzadkowski:2010es}. This example illustrates the main rationale behind the treatment of evanescent operators and demonstrates the computation with both diagrammatic and functional methods. We also discuss the dipole contributions from other evanescent operators arising from leptoquark extensions of the SM model and emphasize the possible ambiguities resulting from the treatment of $\gamma_5$ in these cases. Next, in Section~\ref{sec:evanescent_smeft}, we give prescriptions for the evanescent operators in the SMEFT up to dimension six, and we classify the redundant operators beyond the Warsaw basis needed to account for the evanescent contributions from any tree-level NP mediator. We then present all possible one-loop evanescent SMEFT contributions, which can be readily applied to any BSM study and are provided in the ancillary material. Finally, we discuss the details of two evanescence-free renormalization schemes and demonstrate how to account for evanescent operators in running and matching computations. The methods described here are valid for all orders in the EFT expansion and apply to any EFT. We conclude in Section~\ref{sec:conclusions}. Appendices~\ref{app:ord_eps_shift} and~\ref{app:2-loop_RG} provide further details concerning how to shift the EFT action in the presence of $\ord{\epsilon}$ terms and on two-loop running in the presence of evanescent operators, respectively.

\section{Evanescent Operators: Two Practical Examples}
\label{sec:pheno}
\subsection{Lepton dipoles in the Two-Higgs-Doublet Model}
We begin this section with the 2HDM, where the SM is extended with a second Higgs doublet transforming as $\Phi \sim (\rep{1},\rep{2})_{1/2}$ under the SM gauge group. The Lagrangian of the model reads
\begin{align}\label{eq:2HDMLag}
\mathcal{L}\supset\mathcal{L}_\sscript{SM} + D_\mu \Phi^\dagger D^\mu \Phi - M_\Phi^2\, \Phi^\dagger\Phi - \big( y^{pr}_{\Phi e}\, \overline{\ell}_p \Phi e_r +\mathrm{h.c.} \big) \,,
\end{align}
where we omitted the possible Yukawa couplings with other SM fermions and the rest of the scalar potential, as these terms do not enter in the ensuing discussion. Our conventions for the SM Lagrangian are given in Section~\ref{sec:SM_conventions}.

    \begin{figure}
    \centering
    \raisebox{.45ex}{\includegraphics[scale=.65]{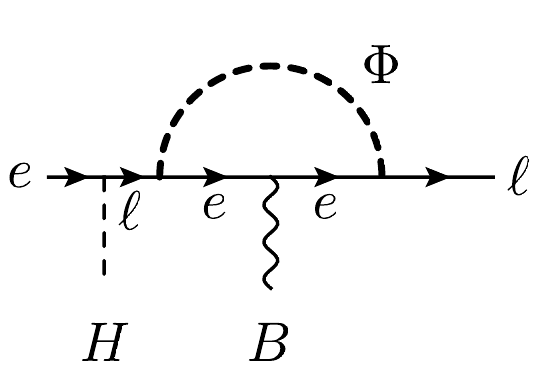}}
    \includegraphics[scale=.65]{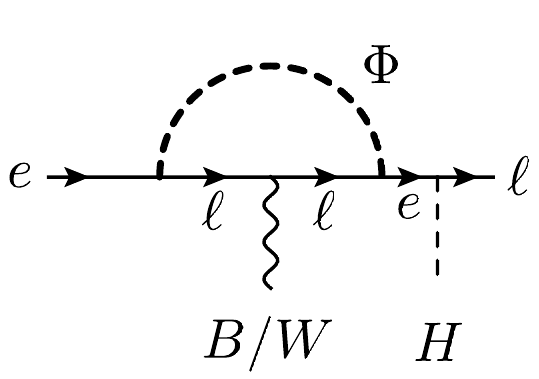}
    \includegraphics[scale=.65]{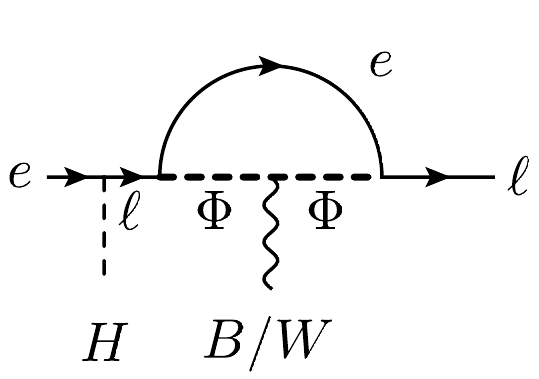}
    \includegraphics[scale=.65]{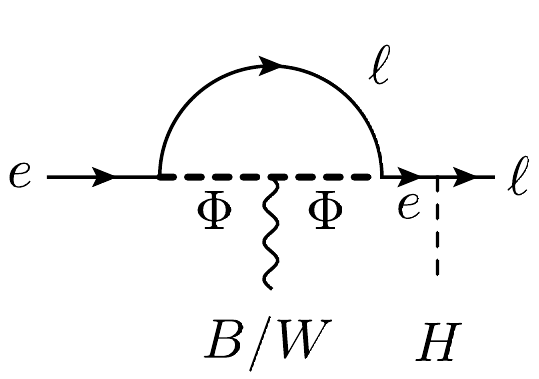}
    \caption{Feynman diagrams generating the dipole part of $e \to \ell B$ and  $e\to\ell W$ amplitudes. Note that the Higgs boson cannot connect to the fermion inside the loop because of hypercharge conservation.}\label{fig:2HDMDipoles}
    \end{figure}

To demonstrate the impact of evanescent operators, we consider the matching to the leptonic dipole operators, which generate lepton flavor--violating processes like $\ell\to\ell^\prime\gamma$ and can place important constraints on the flavor structure of the model. Before we discuss the EFT calculation, we compute the one-loop amplitude using the full model. If we work at fixed order (namely, if we neglect RG resummation), this provides our reference result that the EFT needs to reproduce. Evaluating the Feynman graphs shown in Fig.~\ref{fig:2HDMDipoles} and neglecting irrelevant Dirac structures, we obtain the following amplitudes: 
\begin{align}
    i\mathcal{A}_{e_r \to \ell_p B} &= \frac{-g_Y}{192\pi^2 M_\Phi^2} \! \left[2(2 Y_\ell + Y_\Phi)[y_{e} y_{\Phi e}^{\dagger} y_{\Phi e}]^{pr}   +(2Y_e-Y_\Phi)[y_{\Phi e}y_{\Phi e}^{\dagger} y_{e}]^{pr} \right] \! \big(\bar u \sigma_{\mu\nu} P_R u \big) q^\mu  \varepsilon^{\ast\nu}\,, \nonumber \\
    i\mathcal{A}_{e_r \to \ell_p W} &= \frac{ g_L}{384\pi^2 M_\Phi^2} [ y_{\Phi e}y_{\Phi e}^\dagger y_{e} ]^{pr} \big(\bar u \sigma_{\mu\nu} P_R u\big) q^\mu  \varepsilon^{\ast\nu}\,,
    \label{eq:2HDM_full_loops}
\end{align}
where $y_{e}$ is the lepton Yukawa of the SM Lagrangian, $Y_i$ are the hypercharges, $u$ the lepton spinors, and $ \varepsilon_\mu$ the polarization vector of the gauge boson.  
These expressions can be found in two ways: First, one can directly evaluate the graphs for generic on-shell kinematics and expand them around the limit of large $M_\Phi\gg p_i$, where $p_i$ denotes any external momentum. A simpler way to compute them is to expand the loop integrand around their relevant integration regions for the loop momentum $k$, which in this case means the hard region (where $k \sim M_\Phi\gg p_i$) and the soft region ($M_\Phi\gg k\sim p_i$). Doing so reveals that the amplitude is given by the hard region only. This can easily be seen by noting that the soft-region expansion of the relevant topologies is of the form
\begin{align}
 \left. \int \frac{\dd^dk}{(2\pi)^d}\frac{1}{k^2-M_\Phi^2}\frac{1}{(k+q)^2}\frac{1}{k^2}\right|_\mathrm{soft} = -\frac{1}{M^2_\Phi}\int \frac{\dd^dk}{(2\pi)^d} \frac{1}{(k+q)^2}\frac{1}{k^2} + \ldots \, ,
\end{align}
where $q$ is the momentum of the gauge boson and the dots denote higher powers in the expansion. Since $q^2=0$, the loop is scaleless and the soft region vanishes. As a consequence, the expressions in~(\ref{eq:2HDM_full_loops}) map directly onto the matching coefficients of the dipole operators
\begin{align}
\begin{aligned}
 [Q_{eB}]^{pr} &= \bar\ell_p \sigma_{\mu\nu}e_r\,H B^{\mu\nu}\,, & 
 [Q_{eW}]^{pr} &= \bar\ell_p \sigma_{\mu\nu}e_r\,\tau^I H W^{I\mu\nu}\,,
\end{aligned}
\end{align}
where $\tau^I$ denotes the Pauli matrices. We find that 
\begin{align}\label{eq:CeX_from_matching}
\begin{aligned}
 [C_{eB}]^{pr} &= \frac{g_Y}{384\pi^2 M_\Phi^2} \left[2(2 Y_\ell + Y_\Phi)\, [y_{e} y_{\Phi e}^{\dagger} y_{\Phi e}]^{pr}   +(2Y_e-Y_\Phi)\,[y_{\Phi e}y_{\Phi e}^{\dagger} y_{e}]^{pr} \right]\,,\\
 [C_{eW}]^{pr} &= -\frac{g_L}{768\pi^2 M_\Phi^2} [ y_{\Phi e}y_{\Phi e}^\dagger y_{e} ]^{pr}\,.
\end{aligned}
\end{align}

\begin{figure}
    \centering
    \includegraphics[scale=.7]{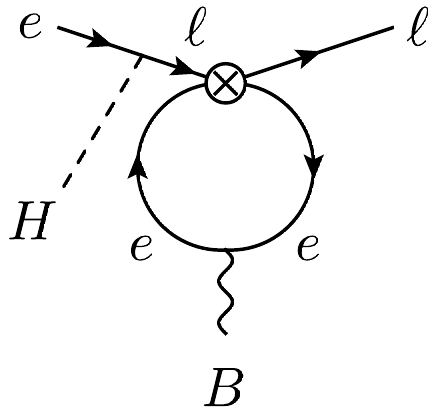}\quad
    \includegraphics[scale=.7]{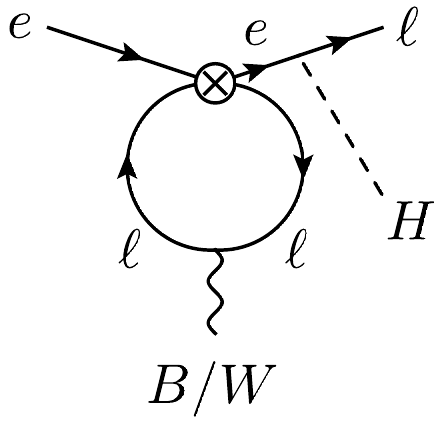}\quad
    \raisebox{1ex}{\includegraphics[scale=.7]{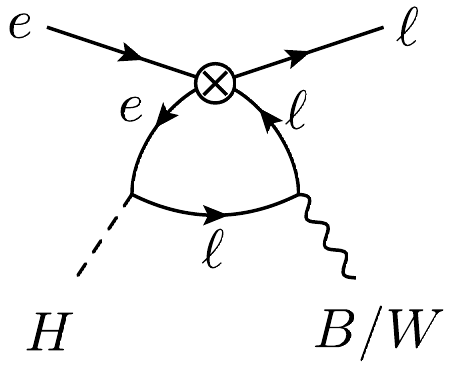}}\quad
    \raisebox{1ex}{\includegraphics[scale=.7]{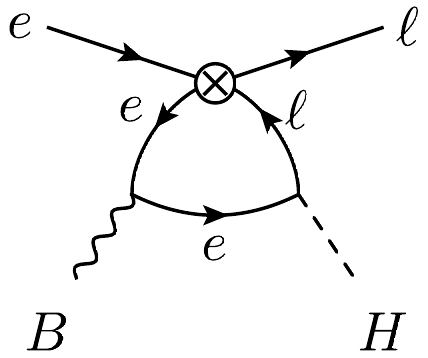}}
    \caption{Feynman diagrams generating dipole matrix elements in the EFT. The crossed circle denotes an insertion of the four-lepton operators found from integrating out the second Higgs doublet $\Phi$ at tree-level.}\label{fig:2HDMEFTLoops}
\end{figure}
Because of the correspondence between the dipole matching coefficients and the full amplitudes, there are no further contributions to the process from one-loop matrix elements with tree-level EFT operators. Integrating out the heavy scalar at tree level yields, among others, the operator and matching coefficient\footnote{In what follows and throughout the manuscript, we denote the Wilson coefficients of the SMEFT operators in the Warsaw basis with the letter `$ C$ ' while we use `$c$' for the ones of the redundant operators.}
\begin{align}
[R_{\ell e}]^{prst} &= (\bar \ell_p e_r)(\bar e_s \ell_t)\,,&
[c_{\ell e}]^{prst} &= \frac{y_{\Phi e}^{pr}y_{\Phi e}^{\ast ts}}{M_\Phi^2}\,. \label{eq:2HDM_treematch}
\end{align}
Inserting this operator into the Feynman graphs shown in Fig.~\ref{fig:2HDMEFTLoops}, one indeed finds that the one-loop amplitude of the EFT vanishes. This can easily be anticipated without a direct computation: The first two graphs are the EFT analogue to the soft region we have computed before. The only scale the loop probes is the virtuality of the photon, which vanishes on-shell. The last two graphs (the second topology), on the other hand, generate only a scalar current, which cannot contribute to the dipole.

We denote the operator with the letter $R$ to show that it is redundant, by which we mean that it is not part of the Warsaw basis~\cite{Grzadkowski:2010es}---the most commonly used SMEFT basis. By means of Fierz identities, it can be transformed into an operator of the Warsaw basis:
\begin{align} \label{eq:2HDMFierz}
 [R_{\ell e}]^{prst} \stackrel{d=4}{=} -\frac{1}{2}(\bar \ell_p \gamma_\mu \ell_t)(\bar e_s \gamma^\mu e_r) \equiv -\frac{1}{2} [Q_{\ell e}]^{ptsr}\,,
\end{align}
which does contribute to the dipole amplitudes. However, this identity has the crucial caveat that it is valid only in $d=4$. In DR with $d=4-2\epsilon$, it is broken starting at $\ord{\epsilon}$. While the difference is irrelevant at tree level, insertions of the two operators into divergent loop integrals yield different results. As we will demonstrate now, these \emph{evanescent} contributions can be absorbed by a finite shift of the action or, equivalently, by the introduction of finite counterterms.

First, we define the evanescent operator as the difference between the redundant and the four-dimensionally reduced object:
\begin{align}
 [E_{\ell e}]^{prst} =[R_{\ell e}]^{prst} + \frac{1}{2}[Q_{\ell e}]^{ptsr}\,.
\end{align}
Next, we check whether this operator produces non-zero loop matrix elements to the dipole transitions.\footnote{We choose to focus on the dipole operators, as these are necessarily loop generated in any UV theory~\cite{Einhorn:2013kja} and, thus, the corrections due to contributions by evanescent operators are potentially of~$\mathcal{O}(1)$. Nevertheless, the evanescent contributions considered here also produce corrections to other Warsaw basis operators not discussed in this example.} This amounts to evaluating the diagrams in Fig.~\ref{fig:2HDMEFTLoops} once more, but this time with insertions of $E_{\ell e}$ instead of $R_{\ell e}$. As the evanescent operator is formally of order~$\mathcal{O}(\epsilon)$, its only relevant contribution (in the four-dimensional limit) stems from the UV poles of divergent loop integrals, and thus it is necessarily local. An economic way to compute this contribution is to expand the loop integrand in the hard region and introduce a fictitious mass $\Lambda$ that also scales as the hard momentum $k\sim \Lambda \gg p_i$. After performing the integration in $k$, one sets $\Lambda\to 0$. Following this prescription, we find the first two diagrams in Fig.~\ref{fig:2HDMEFTLoops} to vanish identically, meaning the term proportional to $R_{\ell e}$ exactly cancels the one proportional to $Q_{\ell e}$. We do, however, find a contribution from the third and fourth diagrams in Fig.~\ref{fig:2HDMEFTLoops}:
\begin{align}
\begin{aligned}
 i\mathcal A^{(E)}_{e\to \ell B} &= -\frac{ig_Y (Y_\ell+Y_e)}{2} [c_{\ell e}]^{prst}y_{e}^{ts} \big(\bar u \sigma_{\mu\nu} P_R u\big) q^\mu  \varepsilon^{\ast\nu} \, \epsilon \int \frac{\dd^dk}{(2\pi)^d}\frac{1}{\left(k^2-\Lambda^2\right)^2}\,, \\
 i\mathcal A^{(E)}_{e\to \ell W} &= -\frac{i g_L}{4} [c_{\ell e}]^{prst}y_{e}^{ts} \big( \bar u \tau^I \sigma_{\mu\nu} P_R u \big) q^\mu  \varepsilon^{\ast\nu}\, \epsilon \int \frac{\dd^dk}{(2\pi)^d}\frac{1}{\left(k^2-\Lambda^2\right)^2}\,.
\end{aligned}
\end{align}
The result is of the form that we anticipated in that the Dirac algebra gives rise to a factor of $\epsilon$, which in turn multiplies a UV-divergent integral. We can compensate these artificial contributions by adjusting the matching coefficients for the dipole operators $Q_{eB}$ and $Q_{eW}$:
\begin{align} \label{eq:CeBCeW_shifts}
 \Delta [C_{eB}]^{pr} &= \frac{3g_Y y_{e}^{ts}}{128\pi^2}[c_{\ell e}]^{prst}\,, &
 \Delta [C_{eW}]^{pr} &= -\frac{g_L y_{e}^{ts}}{128\pi^2}[c_{\ell e}]^{prst}\,.
\end{align}
After supplementing the matching coefficients with these shifts, we can drop $ R_{\ell e}$ in favor of the SMEFT operator $Q_{\ell e}$ following identity~(\ref{eq:2HDMFierz}). The difference in the one-loop matrix elements of the EFT is now correctly accounted for by the shifted matching coefficients.

The above example highlights the importance of correctly handling evanescent operators. The complete basis transformation consists of both a redefinition of the operators as well as their couplings, and performing only one leads to incorrect results. Especially for operators that are generated by the UV theory at the loop level, this amounts to potentially large errors. In our particular example, the basis transformation introduced a one-loop matrix element in the EFT that we knew cannot exist. In the new basis, therefore, we needed to subtract this new contribution by adjusting the matching coefficients, such that the new term was canceled when the EFT matrix element at one-loop is taken. Thus, the basis transformation is: 
\begin{equation}
    \left|
        \begin{array}{rcl}
            [R_{\ell e}]^{prst} &\longrightarrow& -\frac{1}{2} [Q_{\ell e}]^{ptsr}  \\
            \left[ C_{eB,W} \right] ^{pr}  &\longrightarrow & [C_{eB,W}]^{pr} + \Delta [C_{eB,W}]^{pr} \\
            &\vdots&
        \end{array}
    \right.
\end{equation}
where the dots denote the shifts to all operators that we have not considered here, and whose full expression can be found in the supplementary material. Missing the last step would not only have meant making an $\ord{1}$ mistake: Upon inserting the tree-level matching condition~\eqref{eq:2HDM_treematch} in the shifts~\eqref{eq:CeBCeW_shifts}, one finds that the new contributions $\Delta C_{eB,W}^{pr}$ do not even have the same flavor structure as the results in~\eqref{eq:CeX_from_matching}. Specifically, we have:
\begin{align}
 y_{e}^{ts} [c_{\ell e}]^{prst} = \frac{y_{\Phi e}^{pr}}{M_\Phi^2}\, \tr\! \big( y_{e} \,y^\dagger_{\Phi e}\big)\,.
\end{align}
Thus, a consistent treatment of evanescent operators is crucial when matching NP models to the SMEFT and deriving predictions from the results.

\subsection{Dipoles from evanescent operators and reading point ambiguities}
\label{sec:dipole-NDR}

As a second example, we study the effects of evanescent operators from EFT basis transformations among the following four-fermion operators:
\begin{align}\label{eq:Example2Ops}
\begin{aligned}
    [R_{\ell uqe}]^{prst} &= (\bar{\ell}_{ip} u_r) \varepsilon^{ij} (\bar{q}_{js} e_t) \,, 
    &\qquad [R_{u^ce \ell q^c}]^{rtps} &= (\bar{u}^c_r e_t) \epsilon^{ij} (\bar{\ell}_{ip} q^c_{js}) \,,
    \\
    [Q_{\ell equ}^{(1)}]^{ptsr} &= (\bar{\ell}_{ip} e_t) \varepsilon^{ij} (\bar{q}_{js} u_r) \,, 
    & [Q_{\ell equ}^{(3)}]^{ptsr} &= (\bar{\ell}_{ip} \sigma_{\mu\nu} e_t) \varepsilon^{ij} (\bar{q}_{js} \sigma^{\mu\nu} u_r) \,.
\end{aligned}
\end{align}
The redundant operators $R_{\ell uqe}$ and $R_{u^ce \ell q^c}$ are generated at tree level by integrating out the scalar leptoquarks $R_2\sim (\rep{3},\, \rep{2})_{7/6}$ and $S_1\sim (\rep{3},\, \rep{1)}_{\eminus 1/3} $, respectively, while the $ Q_{\ell e qu}^{(1,3)} $ operators are in the Warsaw basis. These operators are related by the Fierz transformation
\begin{align}\label{eq:Fierz-dipole}
    \begin{pmatrix}
    [R_{\ell uqe}]^{prst} \\ [R_{u^ce \ell q^c}]^{rtps}
    \end{pmatrix}
    &=
    \begin{pmatrix}
    -\frac{1}{2} & -\frac{1}{8} \\ -\frac{1}{2} & +\frac{1}{8}
    \end{pmatrix}
    \begin{pmatrix}
    [Q_{\ell equ}^{(1)}]^{ptsr} \\ [Q_{\ell equ}^{(3)}]^{ptsr}
    \end{pmatrix}
    +
    \begin{pmatrix}
    [E_{\ell uqe}]^{prst} \\ [E_{u^ce\ell q^c}]^{rtps}
    \end{pmatrix},
\end{align}
where the four-dimensional identity is supplemented with the addition of the evanescent operators $E_{\ell uqe}^{prst}$ and~$E_{u^ce \ell q^c}^{rtps}$, defined by the difference of                 the operators in the two bases
\begin{align}
\begin{aligned}
[E_{\ell uqe}]^{prst} &\equiv [R_{\ell uqe}]^{prst} +\frac{1}{2} [Q_{\ell e qu}^{(1)}]^{ptsr} +\frac{1}{8} [Q_{\ell e qu}^{(3)}]^{ptsr} \,,\\
[E_{u^ce \ell q^c}]^{rtps} &\equiv [R_{u^ce \ell q^c}]^{rtps} +\frac{1}{2} [Q_{\ell e qu}^{(1)}]^{ptsr} -\frac{1}{8}[Q_{\ell e qu}^{(3)}]^{ptsr} \,.
\end{aligned}
\end{align}
This makes~\eqref{eq:Fierz-dipole} the proper generalization of the four-dimensional Fierz identity to $d=4-2\epsilon$~dimensions. As mentioned, the evanescent operators can be removed in favor of a shift of the Warsaw Wilson coefficients. To derive this shift, we need to compute all UV divergent one-loop diagrams with an insertion of $E_{\ell uqe}$ or $E_{u^ce \ell q^c}$. 

\begin{figure}[t]
    \centering
    \raisebox{0.5ex}{\includegraphics[scale=.7]{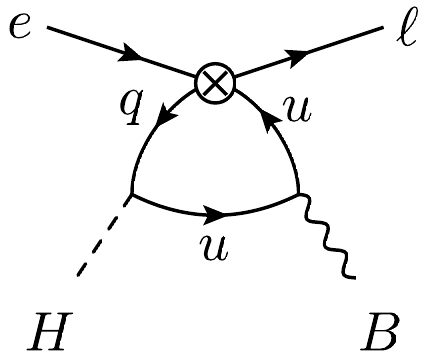}}\qquad\qquad
    {\includegraphics[scale=.7]{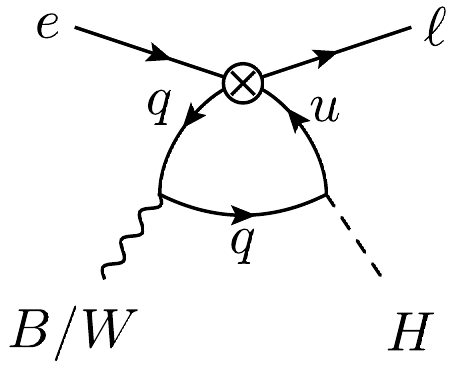}}
    \caption{One-loop diagrams giving evanescent contributions to the dipole operators $Q_{eB}$ and $Q_{eW}$.}
    \label{fig:smeft-dipoles}
\end{figure}

Once again, we focus on the corrections to the lepton dipole operators $Q_{eB}$ and $Q_{eW}$ for concreteness. The relevant Feynman diagrams are shown in Fig.~\ref{fig:smeft-dipoles}, where the crosses denote insertions of the evanescent operators $E_{\ell uqe}$ or $E_{u^ce \ell q^c}$. Alternatively, these contributions can also be computed following the path-integral methods described in~\cite{Cohen:2020fcu,Fuentes-Martin:2020udw}. For the calculation at hand, this amounts to evaluating the following functional supertrace:\footnote{A supertrace is the generalization of the usual operator trace to the case of operators with mixed bosonic and Grassmann components.}  
\begin{align}\label{eq:dip_sTr}
\Delta S_\EFT^{(1)}\supset -i \str \left[ \Delta_q X^\mathrm{E}_{\bar{q}u} \Delta_u X^\mathrm{H}_{\bar{u}q}\right] \Big|_\mathrm{hard}\,,    
\end{align}
Here, the subscript \textit{hard} indicates that we are only interested in the UV poles of the loop integrals. As in the previous example, we extract these UV poles by taking the hard-momntum region defined by a fictitious mass $\Lambda$ that is added to all particle propagators. This way, the functional propagators read $\Delta_{q,u}=1/(i\slashed{D}-\Lambda)$ in position space. The interaction terms are given by
\begin{align}
\begin{aligned}
X^\mathrm{H}_{\bar{u}q} = -\frac{\delta^2\mathcal{L}_\sscript{SM}}{\delta q_{is}\, \delta\bar u_r} 
&= [y_u^*]^{sr}\varepsilon_{ij}\, H^j  P_L\,,\\
X^\mathrm{E}_{\bar{q}u} = -\frac{\delta^2\mathcal{L}_E}{\delta u_r\, \delta\bar{q}_{is}} 
&= [c_{\ell uqe}]^{prst}\,\varepsilon^{ij}\left[|e_t)(\bar\ell_{jp}| +\frac{1}{2}(\bar\ell_{jp} e_t)+\frac{i}{8}(\bar\ell_{jp} \sigma_{\mu\nu} e_t)\,\gamma^\mu\gamma^\nu\right]\\
&\quad+[c_{u^ce\ell q^c}]^{rtps}\,\varepsilon^{ij}\left[|\ell_{jp}^c)(\bar e_t^c| +\frac{1}{2}(\bar\ell_{jp} e_t)-\frac{i}{8}(\bar\ell_{jp} \sigma_{\mu\nu} e_t)\,\gamma^\mu\gamma^\nu\right]\,,
\end{aligned}
\end{align}
with $|\cdot)$ and $(\cdot|$ denoting open fermion lines, $\mathcal{L}_\mathrm{E}= c_{\ell uqe}^{prst}\, E_{\ell uqe}^{prst} + c_{u^ce\ell q^c}^{rtps}\, E_{u^ce\ell q^c}^{rtps}+\mathrm{h.c.}$, and $\mathcal{L}_\sscript{SM}$ as defined in~\eqref{eq:SM_Lagrangian}. 

Applying the Covariant Derivative Expansion (CDE)~\cite{Gaillard:1985uh,Chan:1986jq,Cheyette:1987qz} to the supertrace above, we readily obtain\footnote{After the CDE the functional propagators are effectively reduced to the Feynman propagators, which in momentum space read $\Delta_{q,u}=1/(\slashed{k}-\Lambda)$, with $k$ being the loop momentum.}
\begin{align}
-i\str \left[\Delta_q X^\mathrm{E}_{\bar{q}u} \Delta_u X^\mathrm{H}_{\bar{u}q} \right] \Big|_\mathrm{hard}&\supset \frac{1}{2}\str \left[ \Delta_q \gamma^\mu \Big(g_L \frac{\tau^I}{2} W_{\mu\nu}^I+g_Y Y_q B_{\mu\nu}\Big)\partial_k^\nu\Delta_q X^\mathrm{E}_{\bar{q}u} \Delta_u X^\mathrm{H}_{\bar{u}q} \right] \bigg|_\mathrm{hard} \nonumber\\
&\quad+\frac{1}{2}\str \left[ \Delta_q X^\mathrm{E}_{\bar{q}u} \Delta_u \gamma^\mu g_Y Y_u B_{\mu\nu}\partial_k^\nu \Delta_u X^\mathrm{H}_{\bar{u}q} \right] \Big|_\mathrm{hard}\,.
\end{align}
Each of the supertraces gives contributions that are analogous to the corresponding Feynman diagrams in Fig.~\ref{fig:smeft-dipoles}, with the added benefit of having manifestly covariant expressions. Acting with the momentum derivatives, $ \partial_k$, substituting the $X$ terms, and evaluating the Dirac traces in \emph{naive dimensional regularization}~(NDR) (see Section~\ref{sec:gamma_5} for details), we find\footnote{Throughout this paper we use the convention where $\varepsilon^{0123}=+1$.}
\begin{align}
\begin{aligned}
-i\str \left[ \Delta_q X^\mathrm{E}_{\bar{q}u}\Delta_u X^\mathrm{H}_{\bar{u}q} \right] \Big|_\mathrm{hard}&\supset [y_u^*]^{sr}\left( [c_{\ell uqe}]^{srpt} - [c_{u^ce \ell q^c}]^{rtsp} \right)\int\dd^d x\,(\bar\ell_p \mathcal{F}^{\mu\nu}\sigma^{\rho\sigma} e_t) H\\
&\quad \times\frac{3i}{32}\left[\,2(d-2)\, g_{\mu\rho}g_{\nu\sigma}+(d-6)\, i\varepsilon_{\mu\nu\rho\sigma}\right]\int\frac{\dd^dk}{(2\pi)^d}\frac{1}{\left(k^2-\Lambda^2\right)^2}\\
&\supset\frac{3}{64\pi^2}\,[y_u^*]^{sr}\left( [c_{\ell uqe}]^{srpt} - [c_{u^ce \ell q^c}]^{rtsp} \right)\int\dd^d x\,(\bar\ell_p \mathcal{F}^{\mu\nu}\sigma_{\mu\nu} e_t) H\,,
\end{aligned}
\end{align}
with $\mathcal{F}_{\mu\nu}=g_Y (Y_u+Y_q)B_{\mu\nu} - g_L \tfrac{\tau ^I}{2} W_{\mu\nu}^I$, and where we ignored terms that do not contribute to the dipole operators. In the last line, we have used the four-dimensional identity
\begin{align}\label{eq:EpsRedux}
\varepsilon_{\mu\nu \rho\sigma}\sigma^{\rho \sigma} = - 2 i \sigma_{\mu\nu} \gamma_5\,.
\end{align}
Instead, we could have used any other identity differing from the one above by an $\mathcal{O}(\epsilon)$ term, which would have resulted in a different evanescent contribution. As we discuss in Section~\ref{sec:evanescence_in_EFT}, any such choice for the projection to the four-dimensional basis is valid provided that it is consistently applied in \emph{all} loop computations within the EFT. 

The evaluation of the traces in NDR leads to another subtlety in the present example due to the loss of cyclicity in some of the Dirac traces. The contributions from the redundant operators $R_{\ell uqe}$ and $R_{u^ce \ell q^c}$ do not contain any Dirac traces and can be unambiguously computed in NDR. However, determining the contribution from the Warsaw operator $Q_{\ell e qu}^{(3)}$ requires the computation of a trace over six Dirac matrices and $\gamma_5$, introducing a reading-point ambiguity. The ambiguous traces in this computation are of the form
\begin{align} \label{eq:amb_gamma5_tr}
\begin{aligned}
\tr \big[ \gamma^\alpha \gamma^\rho \gamma^\sigma \gamma_\alpha \gamma^\mu \gamma^\nu \gamma_5 \big]
&= 4i(4-d)\,\varepsilon^{\mu\nu\rho\sigma} \,,\\
\tr \big[ \gamma^\rho \gamma^\sigma \gamma_\alpha \gamma^\mu \gamma^\nu \gamma_5 \gamma^\alpha \big]
&= -4i(4-d)\,\varepsilon^{\mu\nu\rho\sigma} \,. 
\end{aligned}
\end{align}
It is clear that the evaluation of the supertraces depend on where one starts reading the Dirac traces. While we obtained a non-zero dipole contribution for the supertrace~\eqref{eq:dip_sTr}, a vanishing contribution to the dipole operators is found when evaluating instead the supertrace
\begin{align}
\str \left[ \Delta_u X^\mathrm{H}_{\bar{u}q}\Delta_q X^\mathrm{E}_{\bar{q}u}\right]\,,
\end{align}
with an associated change in reading point for the Dirac traces. We find the same results in the diagrammatic approach whenever the reading points are chosen consistently with those in the supertraces. As justified in Section~\ref{sec:gamma_5}, any prescription is valid in the present case as long as it is applied consistently. That means that the reading-point ambiguity in the computation of the one-loop shifts due to evanescent operators has to be the same as the reading point used for the ambiguous one-loop computations within the EFT.

Once we account for the reading-point ambiguity and keep the projection in~\eqref{eq:EpsRedux} for simplicity, we find the following evanescent shift to the dipole operators
\begin{align}\label{eq:dipole-shift}
\Delta \mathcal{L}_\EFT^{(1)} 
&\supset \frac{3}{64\pi^2} \left( 1 - \xi_\mathrm{rp} \right) [y_u^*]^{pr}\left( [c_{\ell uqe}]^{srpt} - [c_{u^ce \ell q^c}]^{rtsp} \right)\left[ g_Y \left( Y_u + Y_q \right)  [Q_{eB}]^{st} -\frac{g_L}{2}[Q_{eW}]^{st} \right],
\end{align}
where we have introduced the parameter $\xi_\mathrm{rp}$ to encode the result dependence on the choice of reading points. This parameter takes the value $\xi_\mathrm{rp}=0$, whenever the Dirac traces are read starting from the Higgs interaction or the propagator coming after it. The value $\xi_\mathrm{rp}=1$, and, thus, $\Delta S_\EFT^{(1)} = 0$, is found for any other reading point. As a result of the accidental cancellation, the dipole calculations in~\cite{Gherardi:2020det,Aebischer:2021uvt} do not need to be corrected as long as any of the reading points yielding $\xi_\mathrm{rp}=1$ and the identity in~\eqref{eq:EpsRedux} are used in the corresponding EFT computations too. For convenience, we recommend ending all Dirac traces at the EFT operators as a consistent reading point also in the SM broken phase.


\section{Evanescence in the SMEFT} \label{sec:evanescent_smeft}
The Warsaw basis~\cite{Grzadkowski:2010es} has become the standard basis for SMEFT studies. It is also the only basis for which the one-loop RG equations~\cite{Jenkins:2013zja,Jenkins:2013wua,Alonso:2013hga} and the one-loop matching to the LEFT~\cite{Dekens:2019ept} are known. Although matching computations often yield EFT operators that are not in the Warsaw basis, it, therefore, becomes desirable to translate them to this basis. The necessary basis translations have to be performed in $d$~dimensions and one needs to account for the evanescent contributions starting at one-loop order. Here, we compute once and for all, the one-loop evanescent contributions resulting from the matching of any weakly-coupled BSM extension to the Warsaw basis. Our result can easily be applied in any (partial or full) one-loop SMEFT matching computation.

\subsection{The SMEFT in the physical basis} \label{sec:SM_conventions}
We begin with the mandatory presentation of our conventions for the SM Lagrangian:
\begin{align}\label{eq:SM_Lagrangian}
\mathcal{L}_\sscript{SM} = 
&-\frac{1}{4} G^A_{\mu\nu} G^{\mu\nu\,A} -\frac{1}{4} W^I_{\mu\nu} W^{\mu\nu\,I} -\frac{1}{4} B_{\mu\nu} B^{\mu\nu} +(D_\mu H)^\dagger (D^\mu H) + \mu^2 (H^\dagger H) - \frac{\lambda}{2} (H^\dagger H)^2 \nonumber\\
&+\sum_{\psi=q,u,d,\ell,e} \bar{\psi}i\slashed{D}\psi -\left(y_u^{pr}\,\bar{q}_p \tilde{H} u_r + y_d^{pr}\,\bar{q}_p H d_r + y_e^{pr}\,\bar{\ell}_p H e_r + \mathrm{h.c.} \right) ,
\end{align}
where we keep the gauge-fixing and ghost Lagrangians implicit. The matter content is defined as per usual, with $p$ and $r$ denoting flavor indices. The conjugated Higgs doublet is given by $\tilde{H}^i = \varepsilon^{ij} H^\ast_j$, where $\varepsilon^{ij}$ is the anti-symmetric $\mathrm{SU}(2)$ tensor defined by~$\varepsilon^{12}=+1$ and $\varepsilon^{ij}=-\varepsilon^{ji}$. 
The covariant derivative is given by
\begin{align}
    D_\mu &= \partial_\mu - i g_s T^A G_\mu^A - i g_L t^I W_\mu^I - i g_Y Y B_\mu \,,
\end{align}
where the $\mathrm{SU}(3)_c$ and $\mathrm{SU}(2)_L$ generators are given by $T^A=\lambda^A/2$ and $t^I=\tau^I/2$, respectively, with $\lambda^A$ being the Gell-Mann matrices and $\tau^I$ the Pauli matrices. For the $\U(1)_Y$ hypercharge,~$Y$, we use the same convention as Ref.~\cite{Grzadkowski:2010es}. The corresponding gauge couplings are dubbed $g_s$,~$g_L$, and~$g_Y$, respectively. The field strength tensors read
\begin{align}
    G_{\mu\nu}^A &= \partial_\mu G_\nu^A - \partial_\nu G_\mu^A + g_s f_{ABC} G_\mu^B G_\nu^C \,, \nonumber\\
    W_{\mu\nu}^I &= \partial_\mu W_\nu^I - \partial_\nu W_\mu^I + g_L \varepsilon_{IJK} W_\mu^J W_\nu^K \,,\\
    B_{\mu\nu} &= \partial_\mu B_\nu - \partial_\nu B_\mu \,, \nonumber
\end{align}
where $f_{ABC}$ and $\varepsilon_{IJK}$ are the $\mathrm{SU}(3)_c$ and $\mathrm{SU}(2)_L$ structure constants, respectively. 

For the SMEFT Lagrangian, we add a tower of higher-dimensional effective operators to the SM Lagrangian~\eqref{eq:SM_Lagrangian}. In this work, we will consider only operators of mass dimension six, for which we adopt the Warsaw basis: 
\begin{align}
    \mathcal{L}_\mathrm{Warsaw} &= \mathcal{L}_\sscript{SM} + \sum_{i} C_i\, Q_i \,, 
\end{align}
with $i$~running over all the dimension-six operators listed in Ref.~\cite{Grzadkowski:2010es}. The dual tensors are defined by $\tilde X^{\mu\nu}=\tfrac{1}{2}\,\varepsilon_{\mu\nu\rho\sigma} X^{\rho\sigma}$, with the $\varepsilon^{0123}=+1$ convention.

\subsection{The SMEFT in \texorpdfstring{$d$}{d}-dimensions and projection to the physical basis}

The Warsaw basis is a physical, on-shell basis of the SMEFT. That is, it is a basis in four spacetime dimensions, as it is constructed using four-dimensional relations like Fierz and Dirac-tensor-reduction identities, as well as field redefinitions, to eliminate redundant operators. However, most loop-level computations in the SMEFT are performed using DR with $d=4-2\epsilon$ space-time dimensions. In $ d \neq 4 $ dimensions, the operators of the Warsaw basis (or any other physical SMEFT basis) do not constitute a proper basis, and new operators have to be added to complete the basis. These redundant operators are related to the ones in the physical basis by the application of intrinsically four-dimensional relations. Their elimination in favor of physical operators gives rise to evanescent operators, which vanish in $ d=4 $~dimensions but generate non-vanishing contributions at higher loop orders. In non-integer dimensions, the basis is infinite-dimensional, but working to a finite order in the loop expansion involves only a finite number of redundant operators. 

In this subsection, we present the complete list of tree-level-generated redundant operators for the Warsaw basis and describe a practical prescription for a physical projection from the $d$-dimensional operator space.\footnote{The method presented here can also be applied to other SMEFT bases.} This implicitly defines all the relevant evanescent structures, which are the subject of this paper.


\subsubsection{Tree-level-generated redundant operators}
\begin{table}[t]
    \renewcommand{\arraystretch}{1.2}
    \centering
    \begin{tabular}{|c|c|c|c|}
        \hline 
        \multicolumn{2}{|c|}{\cellcolor{Gray!25}$(\bar LR)(\bar RL)$ \& $(\bar LR)(\bar LR)$} & \multicolumn{2}{|c|}{\cellcolor{Gray!25}$(\bar R^cR)(\bar RR^c)$}  \\
        \hline
        $R_{\ell e}$   & $(\bar \ell_p e_r)(\bar e_s \ell_t)$              & $R_{e^ce}$         & $(\bar e^c_p e_r)(\bar e_s e^c_t)$                           \\
        $R_{\ell u}$   & $(\bar\ell_p u_r)(\bar u_s \ell_t)$               &  $R_{u^cu}$        & $(\bar u^c_{\alpha p} u_{\beta r})(\bar u_{\beta s} u^c_{\alpha t})$ \\
        $R_{\ell d}$   & $(\bar\ell_p d_r)(\bar d_s \ell_t)$               & $R_{d^cd}$         & $(\bar d^c_{\alpha p} d_{\beta r})(\bar d_{\beta s} d^c_{\alpha t})$ \\
        $R_{qe}$       & $(\bar q_p e_r)(\bar e_s q_t)$                    & $R_{e^c u}$        & $(\bar e^c_p u_r)(\bar u_s e^c_t)$                           \\
        $R_{qu}^{(1)}$ & $(\bar q_p u_r)(\bar u_s q_t)$                    & $R_{e^c d}$        & $(\bar e^c_p d_r)(\bar d_s e^c_t)$                           \\
        $R_{qu}^{(8)}$ & $(\bar q_p T^A u_r)(\bar u_s T^A q_t)$            & $R_{u^c d}$        & $(\bar u^c_{\alpha p} d_{\beta r})(\bar d_{\beta s} u^c_{\alpha t})$ \\
        $R_{qd}^{(1)}$ & $(\bar q_p d_r)(\bar d_s q_t)$                    & $R_{u^c d}^\prime$ & $(\bar u^c_{\alpha p} d_{\beta r})(\bar d_{\alpha s} u^c_{\beta t})$ \\
        $R_{qd}^{(8)}$ & $(\bar q_p T^A d_r)(\bar d_s T^A q_t)$            &                              &                                                      \\ 
        $R_{\ell uqe}$ & $(\bar\ell_{ip} u_r)\varepsilon^{ij}(\bar q_{js} e_t)$  &                              &                                                      \\
        \hline
        \multicolumn{2}{|c|}{\cellcolor{Gray!25}$(\bar L^cL)(\bar LL^c)$}    & \multicolumn{2}{c|}{\cellcolor{Gray!25}$(\bar R^cR)(\bar L L^c)$}                    \\
        \hline
        $R_{\ell^c\ell}$      & $(\bar\ell^c_{ip}\ell_{jr})(\bar\ell_{js}\ell^c_{it})$                     & $R_{u^c d q q^c }$        & $(\bar u^c_{\alpha p} d_{\beta r})\varepsilon^{ij} (\bar q_{\beta is} q^c_{\alpha jt})$ \\           
        $R_{q^cq}$            & $(\bar q^c_{\alpha ip}q_{\beta jr})(\bar q_{\beta js}q^c_{\alpha it})$ &
        $R_{u^c e \ell q^c}$      & $(\bar u^c_p e_r)\varepsilon^{ij}(\bar \ell_{is} q^c_{jt}) $ \\       
        \cline{3-4}
        $R_{q^cq}^\prime$     & $(\bar q^c_{\alpha ip}q_{\beta jr})(\bar q_{\beta is}q^c_{\alpha jt})$ & \multicolumn{2}{c|}{\cellcolor{Gray!25}Baryon number violating} \\   
        \cline{3-4}
        $R_{ q^c \ell}$        & $(\bar q^c_{ip}\ell_{jr})(\bar\ell_{js} q^c_{it})$                         & $R_{q^cqq^c\ell}$ & $\varepsilon_{\alpha\beta\gamma}\varepsilon_{ij}\varepsilon_{kl}(\bar q^c_{\alpha ip} q_{\beta jr} )(\bar q^c_{\gamma ks} \ell_{lt})$\\             
        $R_{q^c \ell }^\prime$ & $(\bar q^c_{ip}\ell_{jr})(\bar\ell_{is} q^c_{jt})$                         & $R_{u^c u d^c e}$ & $\varepsilon_{\alpha\beta\gamma}(\bar u^c_{\alpha p} u_{\beta r})(\bar d^c_{\gamma s} e_t)$ \\ 
        \hline
    \end{tabular}
    
    \caption{Scalar-type four-fermion operators that are redundant in $d=4$ dimensions.}\label{tab:ScaOpList}
\end{table}

Since the aim of this paper is to determine the one-loop contributions from the presence of evanescent operators, we focus on redundant operators that arise from the tree-level exchange of NP particles. We consider all possible tree-level mediators of spin $0$, $1/2$, and $1$ (excluding antisymmetric rank-2 tensors). The most general Lagrangian (with up to dimension-five interactions) for these mediators, and their tree-level matching to the Warsaw basis, can be found in Ref.~\cite{deBlas:2017xtg}. It follows that spin-$1/2$ mediators do not generate redundant operators at tree level, and only operators mediated by NP scalars and vectors need to be considered in our analysis. The complete list of tree-level-generated redundant operators for the Warsaw basis is given in Tables~\ref{tab:ScaOpList} and~\ref{tab:VectOpList}. The list of mediators generating these operators is presented in Table~\ref{tab:mediator-operators}.

\begin{table}[t]
\renewcommand{\arraystretch}{1.2}
\centering
\begin{tabular}{|c|c|c|c|}
    \hline
    \multicolumn{2}{|c|}{\cellcolor{Gray!25}$(\bar LL)(\bar LL)$}                                       & \multicolumn{2}{c|}{\cellcolor{Gray!25}$(\bar L^cR)(\bar RL^c)$}                                         \\
    \hline
    $R_{\ell\ell}^{(3)}$ & $(\bar \ell_p\gamma_\mu\tau^I \ell_r) (\bar \ell_s\gamma^\mu\tau^I\ell_t)$ & $R_{\ell^c e}$     & $(\bar\ell^c_p\gamma_\mu e_r)(\bar e_s\gamma^\mu \ell^c_t)$                        \\
    $R_{qq}^{(1,8)}$     & $(\bar q_p\gamma_\mu T^Aq_r)(\bar q_s\gamma^\mu T^Aq_t)$                   & $R_{\ell^c u}$     & $(\bar\ell^c_p\gamma_\mu u_r)(\bar u_s\gamma^\mu\ell^c_t)$                         \\
    $R_{qq}^{(3,8)}$     & $(\bar q_p\gamma_\mu\tau^I T^Aq_r)(\bar q_s \gamma^\mu\tau^IT^Aq_t)$       & $R_{\ell^c d}$     & $(\bar\ell^c_p\gamma_\mu d_r)(\bar d_s\gamma^\mu\ell_t^c)$                         \\
    $R_{\ell q}^{(1)}$   & $(\bar\ell_p\gamma_\mu q_r)(\bar q_s\gamma^\mu\ell_t)$                     & $R_{q^c e d \ell^c }$ & $(\bar q^c_p \gamma^\mu e_r) (\bar d_s \gamma_\mu \ell^c_t)$                          \\
    $R_{\ell q}^{(3)}$   & $(\bar\ell_p\gamma_\mu\tau^I q_r)(\bar q_s \gamma^\mu\tau^I\ell_t)$        & $R_{q^ce}$         & $(\bar q^c_p\gamma_\mu e_r)(\bar e_s\gamma^\mu q^c_t)$                             \\
    \cline{1-2}
    \multicolumn{2}{|c|}{\cellcolor{Gray!25}$(\bar RR)(\bar RR)$}                                       & $R_{q^cu}$         & $(\bar q^c_{\alpha p} \gamma_\mu u_{\beta r}) (\bar u_{\beta s}\gamma^\mu q^c_{\alpha t})$ \\
    \cline{1-2}
    $R_{uu}^{(8)}$       & $(\bar u_p\gamma_\mu T^Au_r)(\bar u_s\gamma^\mu T^A u_t)$                  & $R_{q^cu}^\prime$  & $(\bar q^c_{\alpha p} \gamma_\mu u_{\beta r}) (\bar u_{\alpha s}\gamma^\mu q^c_{\beta t})$ \\
    $R_{dd}^{(8)}$       & $(\bar d_p\gamma_\mu T^Ad_r)(\bar d_s\gamma^\mu T^A d_t)$                  & $R_{q^cd}$         & $(\bar q^c_{\alpha p} \gamma_\mu d_{\beta r}) (\bar d_{\beta s}\gamma^\mu q^c_{\alpha t})$ \\
    $R_{eu}$             & $(\bar e_p\gamma_\mu u_r)(\bar u_s\gamma^\mu e_t)$                         & $R_{q^cd}^\prime$  & $(\bar q^c_{\alpha p} \gamma_\mu d_{\beta r}) (\bar d_{\alpha s}\gamma^\mu q^c_{\beta t})$ \\
    \cline{3-4}
    $R_{ed}$             & $(\bar e_p\gamma_\mu d_r)(\bar d_s\gamma^\mu e_t)$                         & \multicolumn{2}{c|}{\cellcolor{Gray!25}Baryon number violating}                                          \\
    \cline{3-4}
    $R_{ud}^{(1)}$       & $(\bar u_p\gamma_\mu d_r)(\bar d_s\gamma^\mu u_t)$                         & $R_{d^c\ell q^cu}$ & $\varepsilon_{\alpha\beta\gamma}\varepsilon_{ij} (\bar d^c_{\alpha p}\gamma_\mu\ell_{ir})(\bar q^c_{\beta js}\gamma^\mu u_{\gamma t})$ \\
    $R_{ud}^{(8)}$       & $(\bar u_p\gamma_\mu T^A d_r)(\bar d_s\gamma^\mu T^A u_t)$                 & $R_{u^c\ell q^cd}$ & $\varepsilon_{\alpha\beta\gamma}\varepsilon_{ij} (\bar u^c_{\alpha p}\gamma_\mu\ell_{ir})(\bar q^c_{\beta js}\gamma^\mu d_{\gamma t})$ \\
    \cline{1-2}
    \multicolumn{2}{|c|}{\cellcolor{Gray!25}$(\bar LL)(\bar RR)$}                                       & $R_{q^ceu^cq}$     & $\varepsilon_{\alpha\beta\gamma}\varepsilon_{ij} (\bar q^c_{\alpha ip}\gamma_\mu e_r)(\bar u^c_{\beta s}\gamma^\mu q_{\gamma jt})$ \\
    \cline{1-2}
    $R_{\ell q de}$      & $(\bar\ell_p\gamma_\mu q_r)(\bar d_s\gamma^\mu e_t)$                       &                              &\\
    \hline
\end{tabular}
\caption{Vector-type four-fermion operators that are redundant in $d=4$ dimensions. }\label{tab:VectOpList}
\end{table}

\begin{table}[t]
    \renewcommand{\arraystretch}{1.2}
    \centering
    \resizebox{\textwidth}{!}{
    \begin{tabular}{|c|c||c|c|}
    \hline
    \cellcolor{Gray!25} Scalar field & \cellcolor{Gray!25} Operators & \cellcolor{Gray!25} Scalar field & \cellcolor{Gray!25} Operators\\
    \hline
    $(\rep{1},\rep{1})_0$                    & $-$                            & $(\rep{3},\rep{1})_{\frac23}$         & $R_{d^c d}$ \\
    $(\rep{1},\rep{1})_1$                    & $R_{\ell^c \ell}$    & $(\rep{3},\rep{1})_{\eminus\frac43}$  & $R_{u^c u},R_{e^c d}, {\color{red}R_{u^c u d^c e}} $ \\
    $(\rep{1},\rep{1})_2$                    & $R_{e^c e}$          & $(\rep{3},\rep{2})_{\frac16}$         & $R_{\ell d}$ \\
    $(\rep{1},\rep{2})_{\frac12}$            & $R_{\ell e}, R_{qu}^{(1)}, R_{qd}^{(1)}$ & $(\rep{3},\rep{2})_{\frac76}$        & $R_{\ell u}, R_{qe}, R_{\ell u q e} $\\
    $(\rep{1},\rep{3})_0$                    & $-$                            & $(\rep{3},\rep{3})_{\eminus\frac13}$ & $R_{q^c q},R_{q^c q}^\prime, R_{q^c \ell},R_{q^c \ell}^\prime, {\color{red}R_{q^c q q^c \ell}} $   \\
    $(\rep{1},\rep{3})_1$                    & $R_{\ell^c \ell}$    & $(\rep{6},\rep{1})_{\frac13}$         & $R_{q^cq},R_{q^cq}^\prime,R_{u^c d},R_{u^c d}^\prime,R_{u^c d q q^c }$\\
    $(\rep{1},\rep{4})_{\frac12}$            & $-$                            & $(\rep{6},\rep{1})_{\eminus\frac23}$ & $R_{d^cd}$ \\
    $(\rep{1},\rep{4})_{\frac32}$            & $-$                            & $(\rep{6},\rep{1})_{\frac43}$        & $R_{u^cu}$\\
    $(\rep{3},\rep{1})_{\eminus\frac13}$     & $R_{q^cq},R_{q^cq}^\prime,R_{q^c\ell},R_{q^c\ell}^\prime,R_{e^c u},R_{u^c d},$                                     & $(\rep{6},\rep{3})_{\frac13}$        & $R_{q^c q},R_{q^c q}^\prime$\\
    & $R_{u^c d}^\prime,R_{u^c d q q^c },R_{u^c e \ell q^c},{\color{red}R_{q^cqq^c\ell}}$ & $(\rep{8},\rep{2})_{\frac12}$  & $R_{qu}^{(8)}, R_{qd}^{(8)}$\\[5pt]
    \hline
    \hline
    \cellcolor{Gray!25} Vector field & \cellcolor{Gray!25} Operators & \cellcolor{Gray!25} Vector field & \cellcolor{Gray!25} Operators\\
    \hline
    $(\rep{1},\rep{1})_0$                & $-$                            & $(\rep{3},\rep{2})_{\frac16}$        & $R_{\ell^c u}, R_{q^c d}, R_{q^c d}^\prime, {\color{red}R_{u^c\ell q^cd}}$\\
    $(\rep{1},\rep{1})_1$                & $R_{ud}^{(1)}$       & $(\rep{3},\rep{2})_{\eminus\frac56}$ & $R_{\ell^c d}, R_{q^c e}, R_{q^c u}, R_{q^c u}^\prime, R_{q^c e d \ell^c }, {\color{red}R_{d^c\ell q^cu}}, {\color{red} R_{q^ceu^cq}}$\\
    $(\rep{1},\rep{2})_{\frac12}$        & $-$                            & $(\rep{3},\rep{3})_{\frac23}$        & $R_{\ell q}^{(3)}$\\
    $(\rep{1},\rep{2})_{\eminus\frac32}$ & $R_{\ell^c e}$       & $(\rep{6},\rep{2})_{\eminus\frac16}$ & $R_{q^cd}, R_{q^cd}^\prime$\\
    $(\rep{1},\rep{3})_0$                & $R_{\ell\ell}^{(3)}$ & $(\rep{6},\rep{2})_{\frac56}$        & $R_{q^cu}, R_{q^cu}^\prime$\\
    $(\rep{1},\rep{3})_1$                & $-$                            & $(\rep{8},\rep{1})_0$                & $R_{qq}^{(1,8)}, R_{uu}^{(8)}, R_{dd}^{(8)}$\\
    $(\rep{3},\rep{1})_{\frac23}$        & $R_{\ell q}^{(1)}, R_{ed}, R_{lqde}$ & $(\rep{8},\rep{1})_1$                & $R_{ud}^{(8)}$ \\
    $(\rep{3},\rep{1})_{\frac53}$        & $R_{eu}$             & $(\rep{8},\rep{3})_0$                & $R_{qq}^{(3,8)}$ \\
    \hline
    \end{tabular}
    }
    \caption{Contributions from BSM scalars (top) and vectors (bottom) to the redundant operators. Operators that violate baryon number are highlighted in red.}
    \label{tab:mediator-operators}
\end{table}

\subsubsection{Reduction of Dirac structures}
In four dimensions, the Dirac structures 
	\begin{align}\label{eq:4D_basis}
	\{ \Gamma_i \otimes \tilde\Gamma_i \}\equiv \{ & P_L \otimes P_L, P_R \otimes P_R, P_L \otimes P_R, P_R \otimes P_L ,  \gamma^\mu P_L \otimes \gamma_\mu P_L, \gamma^\mu P_R \otimes \gamma_\mu P_R, \nonumber\\
	& \gamma^\mu P_L \otimes \gamma_\mu P_R, \gamma^\mu P_R \otimes \gamma_\mu P_L , \sigma^{\mu\nu} P_L \otimes \sigma_{\mu\nu} P_L, \sigma^{\mu\nu} P_R \otimes \sigma_{\mu\nu} P_R \}\,,
	\end{align}
constitute a basis for all four-fermion structures. 
These are the only structures appearing in the Warsaw basis. 
In one-loop SMEFT amplitudes, we encounter a number of additional structures in $d$~dimensions, which will have to be reduced to this basis (as a first step) to project onto the physical basis.
The simplest way to handle this reduction is to \textit{define} the evanescent operators via gamma--tensor product decomposition~\cite{Tracas:1982gp,Buras:1989xd,Herrlich:1994kh}: Any bilinear structure $ X_1 \otimes X_2 $ can be expressed in terms of the four-dimensional basis elements~$ \{ \Gamma_i \otimes \tilde\Gamma_i \} $ and an evanescent operator~$ E(X_1,X_2) $, which is explicitly defined by the decomposition using the NDR scheme for $\gamma_5$
	\begin{align}
	X_1\otimes X_2&= \sum_i b_i(X_1,\, X_2)\,\Gamma_i\otimes\tilde\Gamma_i+E(X_1,\, X_2)\,. 
	\end{align} 
Contracting both sides of the above equation with the $ d=4 $~basis elements~$ \{ \Gamma_j \otimes \tilde\Gamma_j \} $,
	\begin{align}\label{eq:EvaOpDef}
	\mathrm{Tr}_d\,(\Gamma_j X_1 \tilde\Gamma_j X_2)&=\sum_i b_i(X_1,\, X_2)\,\mathrm{Tr}_d\,(\Gamma_j \Gamma_i \tilde\Gamma_j \tilde\Gamma_i) + \mathcal{O}(\epsilon^2)\,, \qquad \text{for } j=1,\ldots,10 \,,
	\end{align}
yields a system of equations, which can be solved to find the coefficients~$ b_i(X_1,\, X_2).$\footnote{Throughout the paper, we use the NDR prescription for the $d$-dimensional traces (see Section~\ref{sec:gamma_5}). For the projection to the physical basis through gamma reduction, this does not introduce an ambiguity associated with the reading point of the $\gamma_5$-odd part of the traces because the Lorentz indices are fully contracted, ensuring a vanishing Levi-Civita tensor.} Here, $ \mathrm{Tr}_{d(4)} $ denotes the trace in $ d(4) $~dimensions. Equivalently, we find
	\begin{align}
	E_{\alpha\beta,\gamma \delta}(X_1,X_2) \,(\tilde \Gamma_j)_{\beta\gamma} (\Gamma_j)_{\delta\alpha}&=\mathcal{O}(\epsilon^2)\,,
	\end{align}
where the evanescent structure~$E(X_1,X_2)$ is written in terms of its spinor indices~$\alpha,\beta,\gamma,\delta$.
This implicitly defines evanescent contributions $ E(X_1,\, X_2) $ from the reduction of the Dirac algebra. 
In particular, for the relevant one-loop SMEFT structures at dimension six, we find the relations
\begin{align}\label{eq:EvanescentOperators}
	\gamma^\mu \gamma^\nu P_L \otimes \gamma_\nu \gamma_\mu P_L &= ( 4- 2 \epsilon )\, P_L \otimes P_L + \sigma^{\mu\nu} P_L \otimes \sigma_{\mu\nu} P_L \, , \nonumber\\
	\gamma^\mu \gamma^\nu P_L \otimes \gamma_\nu \gamma_\mu P_R &= 4( 1-2\epsilon )\, P_L \otimes P_R + E_{LR}^{[2]} \, , \nonumber\\
	\gamma^\mu \gamma^\nu \gamma^\lambda P_L \otimes \gamma_\lambda \gamma_\nu \gamma_\mu P_L &= 4(1-2\epsilon )\, \gamma^\mu P_L \otimes \gamma_\mu P_L + E_{LL}^{[3]} \, , \\
	\gamma^\mu \gamma^\nu \gamma^\lambda P_L \otimes \gamma_\lambda \gamma_\nu \gamma_\mu P_R &= 16( 1-\epsilon )\,\gamma^\mu P_L \otimes \gamma_\mu P_R + E_{LR}^{[3]} \, , \nonumber\\
	\gamma^\mu \gamma^\nu \sigma^{\lambda\rho} P_L \otimes \sigma_{\lambda\rho} \gamma_\nu \gamma_\mu P_L &= 16(3-5\epsilon )\,P_L \otimes P_L + 2(6-7\epsilon )\,\sigma^{\mu\nu} P_L \otimes \sigma_{\mu\nu} P_L + E_{LL}^{[4]} \, . \nonumber
\end{align}
Analogous expressions are found for the opposite chirality terms. Here, $\smash{E_{LR}^{[2]}}=6\epsilon\,P_L\otimes P_R+\sigma^{\mu\nu} P_L \otimes \sigma_{\mu\nu} P_R$, which can be obtained from the identity $\gamma_\mu\gamma_\nu=g_{\mu\nu}-i\,\sigma_{\mu\nu}$, while all other evanescent operators are defined directly by~\eqref{eq:EvanescentOperators}. Other choices for the basis $\{ \Gamma_i \otimes \tilde\Gamma_i \}$ and/or $\gamma_5$ schemes yield different definitions for the evanescent operators and different coefficients for the $\epsilon$-terms. In the literature (see e.g.~\cite{Dekens:2019ept}), the $\mathcal{O}(\epsilon)$~coefficients in~\eqref{eq:EvanescentOperators} are often kept generic (often denoted by $a_\mathrm{ev}$,~$b_\mathrm{ev}$,~\ldots) to allow for other definitions of the evanescent operators. Different values for these coefficients require different finite counterterms to compensate the effect of evanescent operators in \emph{evanescence-free} schemes, and thus correspond to different renormalization- or $\gamma_5$-schemes~\cite{Herrlich:1994kh}. In~\eqref{eq:EvanescentOperators}, we fix the $\mathcal{O}(\epsilon)$~coefficients to be consistent with the NDR~scheme. In the ensuing computations, we also use the relation
\begin{align}\label{eq:EpsSigmaRel}
\varepsilon_{\mu\nu \rho\sigma}\sigma^{\rho \sigma} = - 2 i \sigma_{\mu\nu} \gamma_5 + E_{\mu\nu}^{(\varepsilon)}\,,
\end{align}
which defines the evanescent structure $E_{\mu\nu}^{(\varepsilon)}$.

\subsubsection{Fierz transformations}\label{subsec:Fierz}
In the SMEFT, there is additional redundancy among the fermion bilinears due to Fierz relations~\cite{Fierz1937ZurFT}, which relates operators  with rearranged fermion lines. We use the following four-dimensional basis~$\Gamma_n$ for the Dirac algebra~\cite{Nishi:2004st} (and its dual~$\tilde\Gamma_n$):\footnote{This is not a unique basis definition. However, it is a convenient choice when dealing with Warsaw operators.}
	\begin{equation}\label{eq:4DGammaBasis}
	\Gamma_n=\{P_L,P_R,\gamma^\mu P_L,\gamma^\mu P_R,\sigma^{\mu\nu}\}\,,
	\qquad
	\tilde\Gamma_n=\{P_L,P_R,\gamma_\mu P_R,\gamma_\mu P_L,\frac{\sigma_{\mu\nu}}{2}\}\,,
	\end{equation}
which satisfy the orthogonality relation
	\begin{equation}\label{eq:GammaOrtho}
	\mathrm{Tr}_4\,(\Gamma_m\,\tilde\Gamma_n)=2\,\delta_{nm}\,.
	\end{equation}
Also in four dimensions, the fermion bilinears satisfy the Fierz identities
	\begin{equation}\label{eq:Fierz}
	(X_1)\otimes[X_2]=\frac{1}{4}\,\mathrm{Tr}_4\,(\Gamma_n X_1 \tilde\Gamma_m X_2)\,(\tilde\Gamma_n]\otimes[\Gamma_m)\,, \qquad (d= 4)
	\end{equation}
for any Dirac structures $X_1 \otimes X_2$. The resulting identities are needed to map the SMEFT operators to a physical basis. As the Fierz identities~\eqref{eq:Fierz} hold only in four dimensions, the residues from their application in $ d $ dimensions define additional evanescent operators:
	\begin{equation}
	(X_1)\otimes[X_2]=\frac{1}{4}\,\mathrm{Tr}_4\,(\Gamma_n X_1 \tilde\Gamma_m X_2)\,(\tilde\Gamma_n]\otimes[\Gamma_m) + E_\mathrm{Fierz} (X_1,\, X_2) \qquad (d\neq 4). 
	\end{equation}
Together, the gamma reduction and Fierz identities let us cast any two- and four-fermion operator in the Warsaw basis plus an evanescent remnant. This is sufficient to define our prescription for the projection from the $d$-dimensional set of (semi-)fermionic operators to the Warsaw basis.

\subsubsection{Reduction of bosonic structures}

The application of identities involving the Levi-Civita tensor, which is an intrinsically four-dimensional object, can also produce bosonic evanescent operators. For instance, in the SMEFT at mass-dimension six,
the relation
\begin{align}
\varepsilon^{\mu_1 \mu_2 \mu_3 \mu_4} \varepsilon_{\nu_1 \nu_2 \nu_3 \nu_4} = -24\, \delta\ud{\mu_1}{[\nu_1} \delta\ud{\mu_2}{\nu_2} \delta\ud{\mu_3}{\nu_3} \delta\ud{\mu_4}{\nu_4]}\,, 
\end{align}
valid only in $d=4$ dimensions, is required to reduce operators of the form $X \widetilde X^2$ or $\widetilde X^3$, with $X$ ($\widetilde X$) being a (dual) field-strength tensor. Similarly, the Schouten identity
\begin{align}
    0 &= g_{\mu\nu} \varepsilon_{\alpha\beta\gamma\delta} + g_{\mu\alpha} \varepsilon_{\beta\gamma\delta\nu} + g_{\mu\beta} \varepsilon_{\gamma\delta\nu\alpha} + g_{\mu\gamma} \varepsilon_{\delta\nu\alpha\beta} + g_{\mu\delta} \varepsilon_{\nu\alpha\beta\gamma} \,, 
\end{align}
is needed to remove redundant bosonic operators in the SMEFT at dimension eight~\cite{Chala:2021cgt}. As before, the extension of these identities to $d$ dimensions requires the inclusion of new evanescent operators. While none of these identities are necessary for the evanescent contributions computed in this paper, we recommend defining the resulting evanescent operators directly from the four-dimensional identities, as we did with the relation in~\eqref{eq:EpsSigmaRel}.

\subsection{Evanescent contributions to the physical SMEFT Lagrangian} \label{sec:SMEFT_evanescent_computation}
The rules of the last section for reducing operators let us define a projection $ \mathcal{P} $ to bring operators to the physical (Warsaw) basis. With the full set of relevant redundant operators, the evanescent Lagrangian simply reads
\begin{align}\label{eq:EvaLag}
\mathcal{L}_E&= \sum_i c_i \left(R_i-\mathcal{P}\,R_i\right)\,,
\end{align}
where the index $i$ runs over all operators in Tables~\ref{tab:ScaOpList} and~\ref{tab:VectOpList}. In our case, this involves only Fierz identities, following the prescription defined in Subsection~\ref{subsec:Fierz}. In what follows, we describe how to determine the shifts to the Warsaw operators that are needed to remove the evanescent Lagrangian, and present our results at the end. 

\subsubsection{Functional evaluation of the evanescent shifts}

\begin{figure}[t]
    \centering
    \begin{subfigure}[b]{0.12\textwidth}
        \includegraphics[width=\textwidth]{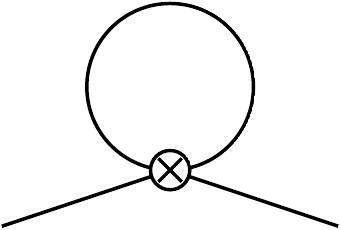}
        \caption{}
    \end{subfigure}
    \hfill
    \begin{subfigure}[b]{0.12\textwidth}
        \includegraphics[width=\textwidth]{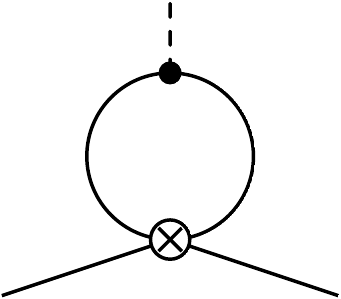}
        \caption{}
    \end{subfigure}
    \hfill
    \begin{subfigure}[b]{0.12\textwidth}
        \includegraphics[width=\textwidth]{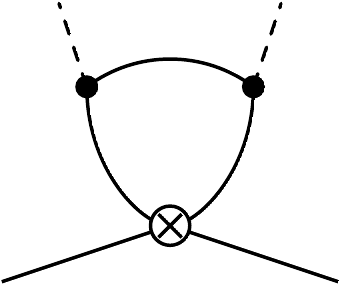}
        \caption{}
    \end{subfigure}
    \hfill
    \begin{subfigure}[b]{0.12\textwidth}
        \includegraphics[width=\textwidth]{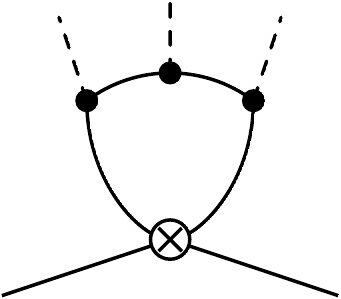}
        \caption{}
    \end{subfigure}
    \hfill
    \begin{subfigure}[b]{0.12\textwidth}
        \includegraphics[width=\textwidth]{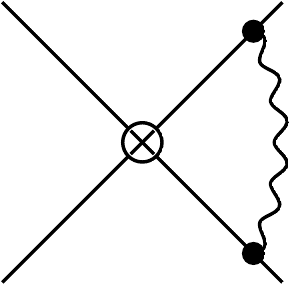}
        \caption{}
    \end{subfigure}
    \hfill
    \begin{subfigure}[b]{0.12\textwidth}
        \includegraphics[width=\textwidth]{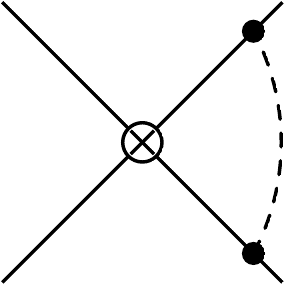}
        \caption{}
    \end{subfigure}
    \caption{Supertrace topologies needed to evaluate the one-loop evanescent contributions to the Warsaw matching. The crossed circles denote evanescent operator interactions whereas black dots denote the SM ones. Solid, wavy, and dashed lines represent fermionic, gauge, and scalar fields, respectively.}
    \label{fig:covLoops}
\end{figure}

The one-loop evanescent shifts to the Warsaw operators are obtained from the computation of all possible one-loop amplitudes with one insertion of the evanescent operators in~\eqref{eq:EvaLag}. Following the approach of~\cite{Fuentes-Martin:2020udw,Cohen:2020fcu}, these are determined from the evaluation of the following power-type supertraces:
\begin{align}\label{eq:EvaSTr}
\Delta S_\EFT^{(1)} &= - \dfrac{i}{2} \sum_{n= 1}^{\infty} \mathcal{P} \str \left[(\Delta X^\sscript{SM})^{n-1} \Delta X^\mathrm{E} \right] \bigg|_\mathrm{hard}\,,
\end{align}
where $\Delta$ and $X$ denote particle propagators and interactions, respectively. In position space, the functional propagators read
\begin{align}\label{eq:Prop}
\Delta_i^{\eminus1}&=\left\{
\begin{array}{ccc}
-D^2-\Lambda^2 & \qquad\quad & \mathrm{(scalar)}\\
i\slashed{D}-\Lambda && \mathrm{(fermion)}\\
-g^{\mu\nu}(-D^2-\Lambda) && \mathrm{(vector)}\\
\end{array}
\right.,
\end{align}
while $X^\mathrm{SM (E)}$ correspond to SM (evanescent) interactions, defined from functional derivatives of the corresponding Lagrangians as
\begin{align}
X_{ij}^\sscript{SM} = \delta^{ij} \Delta^{\! \eminus 1}_i - \frac{\delta^2 \mathcal{L}_\sscript{SM}}{\delta\varphi_j \delta\bar\varphi_i}\,,\qquad X_{ij}^\mathrm{E} = -\frac{\delta^2 \mathcal{L}_\mathrm{E}}{\delta\varphi_j \delta\bar\varphi_i}\,,
\end{align}
with $\varphi_{i,j}$ running over all SM fields.\footnote{In contrast to what we did in the example in Section~\ref{sec:dipole-NDR}, we take complex fields and their conjugates as two independent degrees of freedom. This prescription, which is along the lines of that in~\cite{Fuentes-Martin:2020udw,Cohen:2020fcu,Fuentes-Martin:2016uol}, yields a doubling of the complex fields contributions, explaining the different prefactors between~\eqref{eq:dip_sTr} and~\eqref{eq:EvaSTr}.} As in the example in Section~\ref{sec:dipole-NDR}, the subindex ``hard" in
the supertrace indicates that only the hard part of the loop integrals is relevant to determine the UV poles. These are extracted by adding the fictitious UV mass $\Lambda$ to all particle propagators and taking the hard region defined by this mass.

Since we are working to dimension six in the EFT expansion, the sum in~\eqref{eq:EvaSTr} gets truncated at $n=3$. The only supertraces appearing in this sum are
\begin{align} \label{eq:supertrace-list}
    (\text{a}) &:\ \ \str ( \Delta_\psi X_{\psi\psi}^\mathrm{E} ) \,,
    &
    (\text{b}) &:\ \ \str ( \Delta_\psi X_{\psi\psi}^\sscript{SM} \Delta_\psi X_{\psi\psi}^\mathrm{E} ) \,,
    \nonumber\\
    (\text{c}) &:\ \ \str ( \Delta_\psi X_{\psi\psi}^\sscript{SM} \Delta_\psi X_{\psi\psi}^\sscript{SM} \Delta_\psi X_{\psi\psi}^\mathrm{E} ) \,,
    &
    (\text{d}) &:\ \ \str ( \Delta_\psi X_{\psi\psi}^\sscript{SM} \Delta_\psi X_{\psi\psi}^\sscript{SM} \Delta_\psi X_{\psi\psi}^\sscript{SM} \Delta_\psi X_{\psi\psi}^\mathrm{E} ) \,,
    \\
    (\text{e}) &:\ \ \str ( \Delta_\psi X_{\psi A}^\sscript{SM} \Delta_A X_{A \psi}^\sscript{SM} \Delta_\psi X_{\psi\psi}^\mathrm{E}) \,,
    &
    (\text{f}) &:\ \ \str ( \Delta_\psi X_{\psi\psi}^\sscript{SM} \Delta_\psi X_{\psi\phi}^\sscript{SM} \Delta_\phi X_{\phi\psi}^\mathrm{E} ) \,,
    \nonumber
\end{align}
where $\psi$, $A$, and $\phi$ span the SM fermions, vector bosons, and scalars, respectively. These supertraces can be depicted diagrammatically, in a similar way as with Feynman diagrams, by drawing the $\Delta$ and $X$ terms as propagators and interactions, respectively. We show the diagrammatic depiction for these supertraces in Fig.~\ref{fig:covLoops}, where the letter in front of each supertrace in~\eqref{eq:supertrace-list} corresponds to the letter of the corresponding diagram. These supertraces are evaluated following the procedure described in Ref.~\cite{Fuentes-Martin:2020udw}, and we use the CDE~\cite{Gaillard:1985uh,Chan:1986jq,Cheyette:1987qz} to preserve gauge covariance. This implies that the diagrams in Fig.~\ref{fig:covLoops} should be understood as being dressed by arbitrary gauge boson emissions, with each of these emissions corresponding to higher orders in the CDE. For the computations at hand, only topologies (a)--(c) contain higher-order CDE terms or, equivalently, gauge boson emissions.

\subsubsection{$ \gamma_5$ prescription for $d$-dimensional Dirac traces} \label{sec:gamma_5}
As we calculate the one-loop contribution from the evanescent operators, we inevitably encounter the recurring problem of how to treat $ \gamma_5 $ in DR. This is often cast as a choice between mathematical consistency---using, e.g., the `t Hooft--Veltman scheme~\cite{tHooft:1972tcz,Breitenlohner:1977hr}---or manifest gauge invariance at intermediate stages of the computation. Here we will use the NDR scheme, which is of the latter kind, as this is the standard in most matching computations, and it will make our results compatible with these. It also simplifies many intermediate calculations. 

We define the $ d $-dimensional Dirac algebra to satisfy
    \begin{equation}
    \big\{\gamma_\mu,\, \gamma_\nu \big\} = 2g_{\mu \nu}, \qquad \big\{\gamma_\mu,\, \gamma_5 \big\} = 0, \qquad \gamma_5^2 = 1,
    \end{equation}
where $ \gamma_5 $ is taken to be completely anti-commuting. We further take $ \gamma_5 $ to satisfy the usual four-dimensional identity (with $\varepsilon^{0123}=+1$)
    \begin{equation} \label{eq:gamma5_defining_trace}
    \tr\! \big[\gamma_\mu \gamma_\nu \gamma_\rho \gamma_\sigma \gamma_5 \big] = -4 i\, 
    \varepsilon_{\mu\nu\rho\sigma}\,,
    \end{equation}
while traces with fewer than four (or any odd number of) other $ \gamma $-matrices vanish. This choice for the continuation of the Dirac algebra is known to be inconsistent with the cyclic property of the Dirac trace~\cite{Jegerlehner:2000dz}. The ambiguity does not occur in~\eqref{eq:gamma5_defining_trace}. However, starting with $ \gamma_5 $ in the presence of six other $\gamma$-matrices, there is a loss of cyclicity. That is,
    \begin{equation}
    \tr\! \big[\gamma_{\mu_1} \gamma_{\mu_2} \cdots \gamma_{\mu_{2n}} \gamma_5 \big] = \tr\! \big[\gamma_{\mu_2} \cdots \gamma_{\mu_{2n}} \gamma_5 \gamma_{\mu_1}\big] + \ord{\epsilon}, \qquad \text{for} \quad n \geq 3.
    \end{equation}
This ambiguity is what causes the difference between the two traces in~\eqref{eq:amb_gamma5_tr}.

It was argued in~\cite{Fuentes-Martin:2020udw} that the NDR prescription for $ \gamma_5 $ can be used to achieve unambiguous results for one-loop matching computations. Indeed, renormalizable UV theories do not suffer from any ambiguity at one-loop order. The main challenge comes from the expansion of regions, where new IR divergences result in finite, ambiguous contributions to the matching coefficients. However, as long as the same reading points (whatever they might be) are used to evaluate the corresponding one-loop amplitudes in the EFT, the ambiguities cancel. To be precise, by ``reading point" we refer to which $ \gamma $ matrix is the last in the (non-cyclic) trace.  

For our computation of the evanescent contributions, ambiguities resulting from the NDR prescription can occur only as a result of a fermion loop with six or more regular $ \gamma $-matrices (bearing in mind that all SM fermions are chiral by nature). Of the six covariant topologies in~\eqref{eq:supertrace-list}, only (a)--(d) can result in a Dirac trace, depending on the evanescent operator in question. Each propagator contributes one $ \gamma $-matrix, while the scalar Higgs insertions do not give rise to any. Furthermore, the CDE of the fermion propagators up to $ n $th order in the covariant derivatives can contribute up to $ 2 \lfloor n/2 \rfloor $ $ \gamma $-matrices~\cite{Fuentes-Martin:2020udw}. Diagrammatically, this corresponds to needing a $ \gamma $-matrix from a gauge vertex and the extra fermion propagator associated with it to get a field strength tensor in an amplitude. 

Restricting the computation to dimension six in the SMEFT, we can determine the maximal number of $ \gamma $-matrices that are involved in each of the covariant topologies: the propagators in topology (a) up to third order in the CDE admit up to 3 $ \gamma $-matrices; for (b) up to second order, the number is 4; for (c) to first order there can be 3; and for (d) at leading order, there is up to 4. Thus, the only ambiguous contributions involve two $ \gamma $-matrices in the evanescent operators, which can only come from redundant operators Fierzing to
    \begin{equation}
    Q_{\ell equ}^{(3)prst} = (\overline{\ell}_{ip} \sigma_{\mu\nu} e_r) \varepsilon^{ij} (\overline{q}_{js} \sigma^{\mu\nu} u_t),
    \end{equation}
as it is the only tensor current in the Warsaw basis. Finally, there cannot be any ambiguous $ \gamma_5 $ contributions from the (d) topology, since there are only two external Lorentz indices (from the external tensor bilinear). The Levi--Civita tensor from the $ \gamma_5 $ trace cannot be contracted to a Lorentz scalar in this case without tracing to zero. Accordingly, only the second-order CDE of topology (b) with an evanescent operator involving $ Q_{\ell equ}^{(3)prst} $ can give ambiguous contributions depending on the reading point of the trace. It was remarked already in the one-loop SMEFT-to-LEFT matching computation~\cite{Dekens:2019ept} that the dipole contributions stemming from $ Q_{\ell equ}^{(3)prst} $ are problematic when using a naive Dirac algebra. The authors of~\cite{Dekens:2019ept} made the choice of calculating the contribution from this one operator in the `t~Hooft--Veltman scheme to render it unambiguous, whereas the rest of the computation was performed in NDR. 

To dimension six, SMEFT amplitudes are linear in the EFT operators. As remarked already, there is no ambiguity in evaluating one-loop amplitudes with the redundant operators. We can, thus, decompose the unambiguous amplitude from a generic redundant operator $ R $ as 
    \begin{equation}
    \big\langle R \,\big\rangle_\sscript{SM}^{\!(1)}  = \big\langle R - \mathcal{P} R \,\big\rangle_\sscript{SM}^{\!(1)} + \big\langle \mathcal{P} R \,\big\rangle_\sscript{SM}^{\!(1)},
    \end{equation}
where the $ \langle \, \cdot\,  \rangle_\sscript{SM}^{(1)} $ denotes the one-loop amplitude of the operator in the presence of SM interactions. The first term on the right-hand side is the one-loop amplitude of the evanescent operator associated with $ R $, and the second term the one-loop amplitude from the tree-level operators in the Warsaw basis. Since the left-hand side is unambiguous, we find that it is meaningful to apply our NDR scheme to the contributions on the right-hand side too, as long as the ambiguous Dirac traces are evaluated in the same manner in the evaluation of the evanescent operator as will be used later in the evaluation of one-loop amplitudes in the EFT. This can be done by fixing the same reading point for all traces involving the problematic operators. As mentioned in the example in Section~\ref{sec:dipole-NDR}, we recommend fixing the NDR ambiguity associated with evanescent operators involving $ Q_{\ell equ}^{(3)}$ by choosing the reading point where all Dirac traces end at the dimension-six operator. This choice selects $ \xi_\mathrm{rp} = 1 $ in~\eqref{eq:dipole-shift}.

At this stage, the reader might rightly be worried about whether the NDR treatment used here is compatible with the SMEFT-to-LEFT matching result~\cite{Dekens:2019ept} for loops with $ Q_{\ell equ}^{(3)} $. Since the difference between the `t Hooft--Veltman scheme used in that computation and our prescription is $ \mathcal{O} (\epsilon) $ for the Dirac traces, any discrepancy must be related to the UV pole of the diagram. We have verified that our reading point prescription with $ \xi_\mathrm{rp} = 1 $ happens to give the same result. This means that the two calculations are compatible and can be used directly in the same multi-scale analysis, which greatly simplifies the use of these standard results. It should be emphasized that there was a priori no reason that this should have been the case. It also implies that NDR BSM-to-SMEFT matching computations can be compatible with the SMEFT-to-LEFT matching results.

Finally, we would be remiss if we did not point out that this whole discussion could have been avoided simply by making a minor tweak to the Warsaw basis. Substituting  $ Q_{\ell equ}^{(3)} $ in favor of $ R_{\ell uqe} $, would entirely circumvent the $ \gamma_5 $-related ambiguity in one-loop dimension-six computations. Such a change would make it easier to ensure compatibility between different SMEFT computations, and ease the use of NDR.

\subsubsection{Results for the evanescent contributions in the SMEFT}

The one-loop evanescent contributions from the projection of the operators in Tables~\ref{tab:ScaOpList} and~\ref{tab:VectOpList} onto the Warsaw basis are computed, following the procedure described above, using a modified version of \texttt{Matchete}'s proof of concept~\cite{Matchete}. The resulting expressions, consisting of $d$-dimensional replacement rules for the redundant operators valid at one-loop order, are too lengthy to be shown here. Instead, we opt for providing them as ancillary files. The ancillary material contains a PDF file with these replacement rules, as well as a \texttt{Mathematica} notebook with an interface to navigate around and look up individual operator contributions. The results in \texttt{Mathematica} format can also be used for automating the substitutions in cases when complicated expressions with multiple redundant operators are involved.

\section{Evanescence-Free Renormalization Schemes} \label{sec:evanescence_in_EFT}
Evanescent operators appear in most NLO EFT computations, not only in the SMEFT or LEFT context. Seeing as the literature mainly discusses their treatment in specific applications, we proceed to present a general prescription for handling these operators in generic one-loop matching and two-loop running computations. In particular, the methods described in this section apply beyond the example of evanescence in the SMEFT discussed in Section~\ref{sec:evanescent_smeft}.

\subsection{Matching and projecting to the physical basis}
In matching computations, we generically consider some UV theory $ S_\UV[\Phi,\, \phi] $ with heavy and light degrees of freedom $ \Phi $ and $ \phi $, respectively. At the energy scale set by the masses~$M$ of the heavy states, the UV theory is matched to an appropriate EFT, $S_\EFT[\phi]$, to facilitate computations at energy scales $ E\ll M$. The effective action---the generating functional for all 1PI amplitudes---of the UV theory is reproduced by the EFT effective action up to a certain order in a double expansion in loop order and inverse powers of~$M$. The master formula for off-shell matching of a UV theory to its EFT up to one-loop order reads\footnote{We use the superscript `$ (\ell) $' to denote $ \ell $-loop contributions while the loop-order is given by the $ \hbar $ power, with $ \hbar^0 $ being tree level. The full action is $ S = S^{(0)} + S^{(1)} + \ldots $ and similarly for other perturbative quantities.}
    \begin{equation}\label{eq:master}
    S_\EFT[\phi] = S_\UV^{(0)}[\hat{\Phi}[\phi],\, \phi] +  S_\UV^{(1)}[\hat{\Phi}[\phi],\, \phi] +
    \overline{\Gamma}_\UV^{(1)} \big[\hat{\Phi}[\phi],\, \phi\big] \Big|_\mathrm{hard} + \ord{\hbar^2}\,.
    \end{equation}
Here, the term $\hat{\Phi}[\phi]$ denotes the solution of the heavy-field equations of motion from $ S_\UV^{(0)} $ as a series in $ 1/ M $, and the tree-level EFT action $S_\EFT^{(0)}$ is determined by replacing the heavy fields by this solution in $S_\UV^{(0)}$. For renormalizable UV theories, the one-loop action $S_\UV^{(1)}$ consists exclusively of the counterterms required to renormalize the theory. Conversely, if the UV theory is itself an EFT, $ S_\UV^{(1)} $ can contain additional finite contributions that are one-loop order. Finally, the term `$\overline{\Gamma}^{(1)}_X$' denotes all contributions to the effective action stemming from one-loop diagrams with vertices associated with the tree-level action~$S_X^{(0)}$. The \emph{soft} part\footnote{The method of regions~\cite{Beneke:1997zp,Jantzen:2011nz} describes how loop integrals in DR can be decomposed in momentum regions, which are identified by the singularities of the internal propagators. In the present case, the relevant regions are hard and soft, with the loop momentum $k$ satisfying $k\sim M$ and $k\sim E$, respectively.} of the one-loop diagrams in the UV theory is reproduced one-to-one by one-loop EFT diagrams, meaning that the loop contribution to the EFT action is given by the \emph{hard} part of the UV loops (along with tree-level contributions with vertices from $ S_\UV^{(1)} $)~\cite{Fuentes-Martin:2016uol,Zhang:2016pja}. 

The matching procedure recovers a $ d $-dimensional EFT action. Generally, we cannot expect to recover a four-dimensional operator basis in the matching: simply put, there will be additional irreducible Dirac and Lorentz structures. 
Only by choosing a physical operator basis and defining a projection prescription for extracting the physical part of the EFT operators can we separate out the evanescent contributions. Schematically, this choice defines a decomposition 
	\begin{equation} \label{eq:op_decomposition}
	O_d = \mathcal{P}\, O_d + \mathcal{E_P}\, O_d\,,
	\end{equation}
of the $ d $-dimensional operators $ O_d $, where $ \mathcal{P} $ is the projector onto the physical space and $\mathcal{E_P} \equiv \mathrm{id} - \mathcal{P}$ extracts the evanescent piece, formally of rank $ \epsilon $. This induces a decomposition $ \mathscr{O} = \mathscr{P}_\mathcal{P} \oplus \mathscr{E}_\mathcal{P} $ on the operator space into physical and evanescent parts. Despite the infinite dimensionality of the evanescent space, only a finite number of evanescent operators is needed at a given loop order, making higher-order calculations tractable. It is the choice of $ \mathcal{P} $ that defines the \emph{evanescent prescription} used in a computation.

With the basis choice implicit in the evanescent prescription $\mathcal{P}$, it becomes sensible to discuss renormalization-related questions, such as counterterms. We can define the operator $ \mathcal{K}_{\mathcal{P}} $ that extracts all $1/\epsilon$ poles of UV origin\footnote{This includes all UV poles, as well as all poles that are artificially generated by the application of the method of regions. The infrared poles must be removed by an appropriate definition of physical observables, and not by the introduction of counterterms. Thus, they are not relevant to our discussion here.} in the coefficients of both physical and evanescent basis operators. This assumes that a basis has been chosen also for $\mathscr{E}_\mathcal{P}$, but we leave this unspecified, since the evanescent poles are irrelevant for the purposes of one-loop matching. The pole part satisfies $ \commutator{\mathcal{P}}{ \mathcal{K}_{\mathcal{P}} }= 0 $, which would not generically be the case for $ \commutator{\mathcal{P}'}{ \mathcal{K}_{\mathcal{P}} } $, since transforming one basis to another involves $ \ord{\epsilon} $ terms. For this reason, one must take particular care when subtracting poles for renormalization. Having introduced a definition of the divergent part suitable for the full set of $d$-dimensional operators, we can show that the EFT matching automatically preserves the renormalization of the UV theory:
	\begin{equation} \label{eq:EFT_renormalization}
	\begin{split}
	\mathcal{K}_{\mathcal{P}} S_\EFT^{(1)}[\phi] 
	&= \mathcal{K}_{\mathcal{P}} \! \left[ S_\UV^{(1)}[\hat{\Phi}[\phi],\, \phi] + \overline{\Gamma}^{(1)}_\UV \big[\hat{\Phi}[\phi],\, \phi\big]\Big|_\mathrm{hard} \right] \\
	&= -\mathcal{K}_{\mathcal{P}}  \overline{\Gamma}^{(1)}_\UV \big[\hat{\Phi}[\phi],\, \phi\big] \Big|_\mathrm{soft} \\
	& = - \mathcal{K}_{\mathcal{P}}  \overline{\Gamma}^{(1)}_\EFT[\phi]\,.
	\end{split}
	\end{equation}
The first equality results from the direct application of the master formula in~\eqref{eq:master}, while the second equality follows from proper renormalization of the UV theory and the fact that the sum of hard and soft regions amounts for the full contribution. The last equality results from the identification of the soft part of $ \overline{\Gamma}^{(1)}_\UV $ with the loop contributions generated by tree-level EFT operators. 

We observe that the poles in $ S_\EFT^{(1)} $ automatically cancel the divergent piece of the loops generated by tree-level EFT operators. Thus, $ S_\EFT^{(1)} $ contains the counterterms for the divergent EFT amplitudes and, thus, $ S_\EFT $ constitutes the \textit{bare} EFT action. Any separation of the couplings and counterterms to obtain the EFT in a suitable renormalization scheme requires additional calculations in the EFT---effectively the soft part of the UV theory---to determine the exact quantity to be canceled by the counterterms. This complicates direct matching to EFTs in non-\ms/\msbar schemes, such as ones with finite compensation of the evanescent operators~\cite{Dugan:1990df,Herrlich:1994kh}. 

\subsection{Defining a suitable evanescent renormalization scheme}
\label{sec:matching_to_physical_basis}
Modern matching computations rely on the method of regions to directly identify the \emph{hard} part of the loops in the UV theory as the relevant contribution to the EFT action. No EFT loops or soft loops in the UV theory are computed in the process of the matching. Unfortunately, this prevents matching directly to an evanescent scheme where physical contributions from evanescent operators are exactly compensated by finite counterterms~\cite{Herrlich:1994kh,Dugan:1990df}, as the equality of evanescent loops with local operators belongs to the \emph{soft} part of the effective action. This finitely compensated scheme, which we will refer to as $ C $, constitutes the traditional scheme for handling evanescent operators.  We will obtain a practical prescription for how to bring the EFT to the evanescent scheme and demonstrate that it is physically equivalent to an evanescent subtracted scheme $ S $, where a suitable evanescence-free action is constructed.\footnote{The arguments presented here can also be used to remove $\mathcal{O}(\epsilon) $ terms from the action, as may have been introduced from some choices of physical basis prescription. Alternatively, it can be used for transforming the EFT from one evanescent scheme to another, which is the situation encountered when transforming the EFT from one basis to another. This is described in Appendix~\ref{app:ord_eps_shift}.}

As mentioned, the matching formula~\eqref{eq:master} produces a \emph{bare} $ d $-dimensional EFT action. After having settled on a prescription for the projection to the physical basis $\mathcal{P}$, it is decomposed as 
	\begin{equation} \label{eq:bare_EFT}
	\L_\EFT= \L_\EFT(\bar{g},\, \bar{\eta}) = \mathcal{L}_\mathrm{kin} + \bar{g}_a\, Q^a + \bar{\eta}_i\, E^i\,,
	\end{equation}
with $ Q^a\in \mathscr{P}_\mathcal{P} $ and $ E^i\in \mathscr{E}_\mathcal{P} $ for the physical and evanescent operators, respectively, and similarly for the bare physical and evanescent couplings, $ \bar g_a $ and $ \bar\eta_i $ ($ g $ and $ \eta $ denote the renormalized couplings). As a consequence of the matching, even the renormalized part of the bare couplings will have contributions at all loop orders. At this stage, the $ d $-dimensional effective action of the EFT is given by
	\begin{equation}
	\Gamma_{\EFT} 
	= S^{(0)}_\EFT + S^{(1)}_\EFT+   \overline{\Gamma}^{(1)}_\EFT + \ord{\hbar^2}\,.
	\end{equation}
Having specified a physical basis, only the physical part $ \mathcal{P} \, \Gamma_{\EFT} $ of the effective action is relevant to the four-dimensional limit of amplitudes. To make this observation manifest, and to avoid carrying around cumbersome evanescent operators in all future computations, we look for an EFT action $ S_\EFT^S $ with exclusively physical couplings that reproduces the physical part of the effective action. The action in the $ S $ scheme is parametrized as 
	\begin{equation}
	S^S_\EFT(g) = S_\EFT\big(g+ \delta^S \!g(g),\, \delta^S\! \eta(g) \big)\,,
	\end{equation}
where the counterterms $\delta^S\! g$ and $\delta^S\! \eta$ cancel the UV divergences of the physical and evanescent terms, respectively. The $S$ scheme constitutes an evanescence-free version of \msbar, and its finite part is physical,  namely $\mathcal{P} (1-\mathcal{K}_\mathcal{P}) S^S_\EFT = (1-\mathcal{K}_\mathcal{P}) S^S_\EFT$. The corresponding effective action is 
	\begin{equation}
	\Gamma_\EFT^S = S_\EFT^{S (0)} + S_\EFT^{S (1)} + \overline{\Gamma}^{S(1)}_\EFT   + \mathcal{O}(\hbar^2)\,.
	\end{equation}

Evidently, $S_\EFT$ and $S_\EFT^S$ are not identical. However, we require that the EFT in the physical basis reproduces the physical projection of the originally matched EFT action $S_\EFT$. Namely, we enforce the requirement
	\begin{equation} \label{eq:physical_equivalent_matching}
	\mathcal{P}\, \Gamma^S_\EFT = \mathcal{P}\, \Gamma_\EFT\,.
	\end{equation} 
At tree level, we recover the expected relation 
	\begin{equation} \label{eq:SP_1-tree}
	S_\EFT^{S (0)} = \mathcal{P}\, S_\EFT^{ (0)}\,,
	\end{equation}
meaning that a simple projection of the tree-level result is sufficient to get the tree-level part of the action made of physical operators. It is at the one-loop level that we encounter the complications at the center of the issue. Condition~\eqref{eq:physical_equivalent_matching} results in  
	\begin{equation} \label{eq:SP_1-loop}
	\mathcal{P}\, S_\EFT^{S (1)} = \mathcal{P}\, S_\EFT^{(1)} + \Delta^{\!S}\! S_\EFT^{(1)}\,,\qquad  
	\Delta^{\!S}\! S_\EFT^{(1)} \equiv \mathcal{P}\! \left( \overline{\Gamma}^{(1)}_\EFT - \overline{\Gamma}^{S(1)}_\EFT  \right)\,.
	\end{equation}
The term $ \Delta^{\!S}\! S_\EFT^{(1)} $ is the difference between all one-loop contributions with vertices from $ S_\EFT^{(0)} $ and those same loops with vertices from $ S_\EFT^{S (0)} $. In other words, these are all the loops with at least one evanescent operator, which we removed from the tree-level $ S $-scheme EFT action. The evanescent operators have rank $ \mathcal{O}(\epsilon) $, ensuring that their finite contribution to the physical part of the effective action can be traced to the UV divergent piece of the loop integrals. The locality of the loop divergence ensures that $ \Delta^{\!S}\! S_\EFT^{(1)} $ is local. The physical contributions from the evanescent operators are local at higher-loop orders too, which follows from an inductive argument due to Dugan and Grinstein~\cite{Dugan:1990df}: by suitable renormalization of all subdivergences of amplitudes with evanescent contributions, physical contributions can be obtained only from the overall divergence of the diagrams, ensuring locality.    

It is worth pointing out that the counterterms in $ S_\EFT^{S (1)} $ automatically renormalize $ \Gamma_\EFT^S $, on the physical space defined by $ \mathcal{P} $. From the finiteness of $ \Delta^{\!S}\! S_\EFT^{(1)} $ and the renormalization of the full EFT~\eqref{eq:EFT_renormalization}, it follows that
	\begin{equation}\label{eq:one_loop_S_action}
	\mathcal{P} \, \mathcal{K}_{\mathcal{P}} S_\EFT^{S (1)} = \mathcal{P} \, \mathcal{K}_{\mathcal{P}} S_\EFT^{(1)}
	=- \mathcal{P} \,\mathcal{K}_{\mathcal{P}} \overline{\Gamma}^{(1)}_\EFT  
	=- \mathcal{P}\,\mathcal{K}_{\mathcal{P}} \overline{\Gamma}^{S(1)}_\EFT\,.
	\end{equation}
The evanescent part of $ S_\EFT^{S(1)} $ consists of divergent counterterms and does not appear in the computation of physical amplitudes. At NLO, the counterterms $ \delta^S\! \eta $ are relevant only in the two-loop running of the physical couplings in the subtracted $ S $ scheme, as we will see in the next section. 

The subtracted $ S $ scheme defined above is closely related to the finitely compensated $ C $ scheme for the evanescent operators, where their contributions to physical amplitudes are compensated by finite counterterms. Contrary to the $ S $ scheme, the EFT action in the $ C $ scheme equals the bare action (in this manner, it is a more conventional renormalization scheme):
	\begin{equation}
	S_\EFT(\bar{g},\, \bar{\eta}) = S_\EFT^C(g,\, \eta) \equiv S_\EFT\big(g+ \delta^C g + \Delta^{\!C} g,\, \eta + \delta^C \eta \big)\,,
	\end{equation}
where the counterterms $ \delta^C g(g, \eta) $, $ \Delta^{\!C} g(g, \eta) $, and $ \delta^C \eta(g, \eta) $ depend on the renormalized couplings. The particular feature of the $ C $ scheme is the introduction of finite counterterms~$ \Delta^{\!C} g $ for the physical couplings that compensate the loop contributions involving evanescent operators. That is, we define 
	\begin{equation} 
	S_\EFT(\Delta^{\!C} g^{(1)},\, 0) \equiv 
	\mathcal{P} \Big[
	\overline{\Gamma}_\EFT^{(1)} (g^{(0)},\, 0)
	- \overline{\Gamma}_\EFT^{(1)} (g^{(0)},\, \eta^{(0)})
	 \Big]
	= - \Delta^{\!S}\! S_\EFT^{(1)}\,.
	\end{equation}  
The $ \delta^C g $ and $ \delta^C \eta $ counterterms are just the regular \msbar counterterms, required to cancel the UV divergences of the theory. Equating the actions $ S_\EFT^C(g,\, \eta) = S_\EFT(\bar{g},\, \bar{\eta}) $ then yields\footnote{Recall that the renormalized couplings have nontrivial one-loop contributions from the matching of the UV theory to the EFT.}
	\begin{align} \label{eq:R_scheme_EFT_couplings}
	\begin{aligned}
	g^{(0)} &= \bar{g}^{(0)}\,, &\qquad\qquad
	\eta^{(0)} &= \bar{\eta}^{(0)}\,,  \\
	g^{(1)} &=  (1 - \mathcal{K}_\mathcal{P}) \bar{g}^{ (1)} - \Delta^{\! C} g^{(1)}\,, & 
	\eta^{(1)} &= (1 - \mathcal{K}_\mathcal{P}) \bar{\eta}^{(1)}\,, \\
	\delta^C g^{(1)} &= \mathcal{K}_\mathcal{P} \bar{g}^{(1)}\,, &
	\delta^C \eta^{(1)} &= \mathcal{K}_\mathcal{P} \bar{\eta}^{(1)}\,.  
	\end{aligned}
	\end{align}
Again, computing the physical part of the effective action in the $ C $ scheme, 
	\begin{equation}
	\begin{split}
	\mathcal{P}\,\Gamma_\EFT^{C} &= \mathcal{P}\, S^{C\,(0)}_\EFT + \mathcal{P}\, S^{C\,(1)}_\EFT +  \mathcal{P}\, \overline{\Gamma}_\EFT^{C(1)}  + \ord{\hbar^2}\\
	&= S_\EFT\big(g+ \delta^C g,\, 0 \big) +  \mathcal{P}\, \overline{\Gamma}_\EFT^{(1)} (\bar g^{(0)},\, 0) + \ord{\hbar^2}\,,
	\end{split}
	\end{equation}
it is evident that all contributions from the evanescent couplings drop out. The physical part of the effective action $\Gamma_\EFT^C$ does not dependent on the renormalized $ \eta $ and agrees with $ \Gamma^S_\EFT $ by construction. It follows from comparing~\eqref{eq:SP_1-loop} and~\eqref{eq:R_scheme_EFT_couplings} that the renormalized physical couplings are identical in $ S $ and $ C $ schemes.  
We have
	\begin{equation}
	S_\EFT^{S}(g) = S_\EFT^{C}(g,\, 0) \sim_\mathcal{P} S_\EFT^{C}(g,\, \eta) = S_\EFT(\bar{g},\, \bar{\eta})\,, 
	\end{equation}
where `$ \sim_\mathcal{P} $' indicates that the two actions reproduce the same physical effective action. The intuition here is clear: the finite renormalization of the evanescent operators in the $ C $ scheme, ensures that they have no physical effect. Hence, the same physics is produced by simply disregarding the evanescent operators and their counterterms. Indeed, we will proceed to show that the RG flow of the physical couplings also agrees between both schemes.  

Before discussing the running, let us look at what this all means for the SMEFT. The one-loop shift of the renormalized coefficients in~\eqref{eq:EvaSTr} due to the presence of evanescent operators is identified with the shift $ \Delta^{\!S} \! S_\EFT^{(1)} $, whose functional expression is generalized to 
	\begin{equation}
	\begin{split}
	\Delta^{\!S} \! S_\EFT^{(1)} 
	&= - \dfrac{i}{2} \sum_{n= 1}^{\infty} \dfrac{1}{n} \mathcal{P} \str \left[ (\Delta X_\EFT)^n - (\Delta X^{S}_\EFT)^n \right] \bigg|_\mathrm{hard}\,,
	\end{split}
	\end{equation}
which holds in any EFT up to arbitrary order in the mass expansion. Hence, the replacement rules for the redundant SMEFT operators discussed in Section~\ref{sec:SMEFT_evanescent_computation} let us determine the dimension-six SMEFT action in either $ S $ or $ C $ schemes.

\subsection{Physical two-loop running}
\label{sec:P-scheme_running}
The evanescent operators influence the two-loop running of the EFT even when their coefficients have been set to zero. Indeed, the physical operators can generically flow into evanescent operators, which then feed back into the physical coefficients again. It was observed in~\cite{Dugan:1990df,Herrlich:1994kh} that by finitely compensating the evanescent couplings, the running of the physical couplings becomes independent of them. Here we derive the \befs for the physical couplings in the $ S $ scheme, which turns out to be much simpler to do than in the $ C $ scheme, and we show that they are independent of the evanescent couplings. We further demonstrate in Appendix~\ref{app:2-loop_RG} that the \befs are identical to the ones of the $C$ scheme, cementing the equivalence between them.

We consider here a generic EFT in the $ d $-dimensional operator basis, like in~\eqref{eq:bare_EFT}, equipped with a physical prescription $ \mathcal{P} $ and denote the collective set of couplings $ \lambda_{I} = (g_a,\, \eta_i) $. In renormalization schemes such as $ S $ and \msbar without finite counterterms, the \befs of the couplings are given by 
    \begin{equation}
    \beta_I = \dfrac{\dd \lambda_I}{\dd t} =  2 \sum_{\ell = 1}^\infty \ell\, \delta \lambda_{I,1}^{(\ell)}\,,
    \end{equation}
where $ \lambda_{I,1}^{(\ell)}(\lambda) $ is the $ \ell $-loop contribution to the coefficient of the simple $ \epsilon $ pole of the associated counterterm (more details are provided in Appendix~\ref{app:RG_functions}). 
We can then consider the flow of the Lagrangian $\L^S_\EFT(g,\, t)$ in the subtracted $ S $ scheme of Section~\ref{sec:matching_to_physical_basis}, denoting the explicit scale dependence by $ t $. The change in couplings under an infinitesimal change in RG scale reads 
	\begin{equation}
	\begin{split}
	\L^S_\EFT(g,\, t) &= \L^{\overline{\sscript{MS}}}_\EFT(g,\, \eta= 0,\,  t) = \L^{\overline{\sscript{MS}}}_\EFT(g + \delta t\, \beta_g ,\, \delta t \,\beta_\eta,\,  t + \delta t) \\
	&\sim_\mathcal{P} \L^S_\EFT \!\big(g + \delta t( \beta_g + \beta_\eta K),\, t + \delta t\big)\,,
	\end{split}
	\end{equation}
where $K\ud{i}{a}$ denotes the linearized shift in physical coefficient associated with the removal of evanescent operators from the EFT. At one-loop order it is given by
	\begin{equation}
	\int \dd^d x\, K\ud{i}{a}(g) Q^a \equiv  \dfrac{\partial}{\partial \eta_i} \Delta^{\!S}\! S_\EFT \bigg|_{\eta=0} 
	= \mathcal{P} \dfrac{\partial}{\partial \eta_i}
	\overline{\Gamma}_\EFT^{(1)} \bigg|_{\eta=0} + \ord{\hbar^2}\,,
	\end{equation}
as follows from~\eqref{eq:SP_1-loop}. 

Although the RG flow tends to run the Lagrangian back into evanescent space, we observe that a shift is sufficient to bring it back to a form compatible with the $ S $ scheme at a cost of modifying the naive \befs. 
All told, the two-loop \befs of the physical couplings in the $ S $ scheme are given by
	\begin{equation} \label{eq:2-loop_RG}
	\beta^S_a(g) = 2 \delta^S\! g_{a,1}^{(1)} + 4 \delta^S\! g_{a,1}^{(2)} + 2 \delta^S\! \eta_{i,1}^{(1)} K\ud{(1)\, i}{a} + \ord{\hbar^3}\,,
	\end{equation}
and is independent of the evanescent couplings. This expression successfully reproduces the NLO QCD contribution to the anomalous dimension of the four-quark operators from~\cite{Buras:2000if}. Even with a formally infinite operator space, the expression above can be directly applied since only a finite number of evanescent counterterms are needed at a given order in the double expansion in loops and operator dimension.

\section{Conclusions}\label{sec:conclusions}

EFTs play an ever-prominent role as versatile tools to explore physics beyond the SM. Despite this, the subject of evanescent operators, present in any NLO EFT analysis, remains somewhat nebulous in parts of the literature, especially outside the realm of the LEFT. We hope that the examples and general discussion we provide here clarify the importance of a consistent treatment of evanescent operators for BSM studies beyond tree level. Indeed, evanescent operators provide relevant contributions, which are typically of the same size as the corresponding matching corrections for observables generated at the loop level.

In this paper, we analyzed the evanescent contributions to the SMEFT in the Warsaw basis. Only a finite set of redundant operators generated by the tree-level exchange of new heavy scalars and vectors gives rise to evanescent contributions at one-loop order. Consequently, we were able to compute \emph{all} possible BSM evanescent contributions using functional techniques. Our results are provided in the ancillary material in \texttt{Mathematica} and PDF formats, allowing anyone to easily account for evanescent contributions when matching UV theories to the SMEFT at one-loop order.

We appreciate that the ongoing challenge of developing comprehensive and reliable matching frameworks for BSM physics extends beyond the SMEFT as the target EFT. Accordingly, we have described a generic scheme for handling evanescent operators in EFTs, generalizing the understanding from flavor physics calculations. This provides a systematic approach to handling evanescent operators and computing their contribution in generic EFT matching, which will be useful in a variety of BSM computations. We have also shown explicitly that the usual approach of compensating evanescent couplings with finite counterterms can just as well be understood as completely removing the evanescent operators from the theory. This perspective is particularly helpful when determining the influence of evanescent operators in \bef computations.  

This paper paves the way for the automation of EFT basis reduction and translation beyond tree level, an essential element in any automated matching (and running) program. In particular, the results we present here can readily be used to complete the substitution rules employed in \texttt{matchmakereft}~\cite{Carmona:2021xtq} to reduce SMEFT operators to the Warsaw basis. A fully automated handling of evanescent operators in arbitrary EFT constructions, necessary for the treatment of more general matching scenarios, is left for future work.

\subsection*{Acknowledgements}
The authors are grateful to the Mainz Institute for Theoretical Physics (MITP) of the Cluster of Excellence PRISMA$^+$ (Project ID 39083149) for its hospitality and support. JF thanks Matthias Neubert, and the Theoretical High Energy Physics Department at JGU Mainz for their hospitality and support during his stay as a visitor. The work of JF has been supported by the Spanish Ministry of Science and Innovation (MCIN) and the European Union NextGenerationEU/PRTR under grant IJC2020-043549-I, by the MCIN grant PID2019-106087GB-C22, and by the Junta de Andaluc\'ia grants P21\_00199, P18-FR-4314 (FEDER) and FQM101. The research of MK is supported in part by the Deutsche Forschungsgemeinschaft (DFG, German Research Foundation) through the Sino-German Collaborative Research Center TRR110 “Symmetries and the Emergence of Structure in QCD” (DFG Project-ID 196253076, NSFC Grant No. 12070131001, - TRR 110). JP and FW received funding from the European Research Council (ERC) under the European Union's Horizon 2020 research and innovation programme under grant agreement 833280 (FLAY), and by the Swiss National Science Foundation (SNF) under contract 200020-204428. The work of JP is supported in part by the U.S.\ Department of Energy (DOE) under award number~DE-SC0009919. The work of AET has received funding from the Swiss National Science Foundation (SNF) through the Eccellenza Professorial Fellowship ``Flavor Physics at the High Energy Frontier'' project number 186866.

\renewcommand{\thesection}{\Alph{section}}
\appendix

\section{Shifting the Tree-Level EFT Action} \label{app:ord_eps_shift}
Depending on the evanescent prescription, the projection operator $ \mathcal{P} $ may introduce new $ \ord{\epsilon} $ coefficients multiplying the physical operators. These extra terms give rise to finite shifts in the effective action when picking out the pole part of loop diagrams. To remove this extra part from the tree-level EFT, we simply consider a suitably shifted theory 
	\begin{equation}
	\tilde{S}_\EFT^{S (0)} = S_\EFT^{S (0)} + \ord{\epsilon}\,.
	\end{equation}
The requirement that $ \mathcal{P}\, \tilde{\Gamma}^S_\EFT = \mathcal{P}\, \Gamma^S_\EFT + \ord{\epsilon} $, ensuring correct reproduction of the physics, dictates a finite shift in the one-loop EFT action:
	\begin{equation} \label{eq:StP_1-loop}
	\tilde{S}_\EFT^{S (1)} = \mathcal{P}\, S_\EFT^{(1)} + \tilde{\Delta}^{\!S}\! S_\EFT^{(1)}\,, \qquad  
	\tilde{\Delta}^{\!S}\! S_\EFT^{(1)} = \mathcal{P} \Big( \overline{\Gamma}^{(1)}_\EFT - \widetilde{\overline{\Gamma}}^{S(1)}_\EFT  \Big)\,.
	\end{equation}
Regardless of the evanescent prescription $ \mathcal{P} $, we can pick $ \tilde{S}_\EFT^{S\, (0)} $ as the result of using four-dimensional identities to match to the basis of physical operators. This can be compensated for with the shift $ \tilde{\Delta}^{\!S}\! S_\EFT^{(1)} $ in the one-loop action.

The importance of applying the evanescent prescription consistently at all stages along the RG flow is often emphasized. If we consider a second evanescent prescription $ \mathcal{P}' $, we can easily derive the finite change in the one-loop action as 
	\begin{equation}
	S_\EFT^{S'\, (1)} = \mathcal{P}'\, S_\EFT^{S(1)} + 
	\mathrm{P'}\! \left( \overline{\Gamma}^{S(1)}_\EFT - \overline{\Gamma}^{S'(1)}_\EFT  \right)\,,
	\end{equation}
which ensures that the effective action agrees on the physical space defined by $ \mathcal{P}' $. The first term, $ \mathcal{P}'\, S_\EFT^{S(1)} $, is somewhat subtle. The projection $ \mathcal{P}' $ can potentially extract finite pieces from the old counterterms of $  S_\EFT^{S(1)} $. More generally $ (1- \mathcal{K}_{\mathcal{P}'}) \mathcal{K}_{\mathcal{P}} \neq 0 $. This is particularly relevant when the counterterms of an EFT are not explicitly provided. In terms of matching computations, this is also the reason why one must be careful when dropping $ \epsilon $ poles from the computation when implicit \msbar renormalization is assumed. More precisely, the poles can only be dropped after the operators have been decomposed in terms of the relevant operator basis.

\section{Two-Loop Running with Evanescent Operators}
\label{app:2-loop_RG}
We show in detail that the \befs of the physical couplings are independent of the evanescent couplings in the $ C $ scheme. This generalizes the arguments of~\cite{Herrlich:1994kh} beyond four-fermion operators.

\subsection{Renormalization group functions} \label{app:RG_functions}
First, we derive generic RG formulas for the renormalized couplings following standard methods. Let us consider a theory $ \L(\bar{\lambda}) $ in a generic renormalization scheme:
	\begin{equation}
	\bar{\lambda}_{I} = \mu^{k_I \epsilon} (\lambda_I + \delta \lambda_I), \qquad \delta \lambda_I = \sum_{n= 0}^{\infty} \dfrac{\delta \lambda_{I,n}}{\epsilon^{n}},
	\end{equation}
allowing for finite counterterms $ \delta \lambda_{I,0}$.
As the bare coupling is an RG invariant, it follows that 
	\begin{equation} \label{eq:bare_coupling_invariance}
	\begin{split}
	0 =\, & \dfrac{\dd \bar{\lambda}_{I}}{\dd t} = 
	 \epsilon \,k_I (\lambda_I + \delta \lambda_I) + \hat{\beta}_I + \hat{\beta}_J \partial^J \delta \lambda_I \\
	=\, & \epsilon \big[k_I(\lambda_I + \delta \lambda_{I,0}) + \beta'_I + \beta_J' \partial^J \delta \lambda_{I,0} \big] 
	+ \big[k_I \delta \lambda_{I,1} + \beta_I + \beta_J' \partial^J \delta \lambda_{I,1} + \beta_J \partial^J \delta \lambda_{I,0} \big]\\
	& + \sum_{n=1}^{\infty} \dfrac{1}{\epsilon^n} \big[ k_I \delta \lambda_{I,n+1} + \beta_J' \partial^J \delta \lambda_{I,n+1} + \beta_J \partial^J \delta \lambda_{I,n} \big]\,,
	\end{split}
	\end{equation}
where $ \hat{\beta}_I = \partial_t \lambda_{I} = \epsilon\,  \beta'_I + \beta_I $. 
From the first term, we have
	\begin{equation}
	\begin{split}
	\beta_I' &= - k_I(\lambda_{I} +\delta \lambda_{I,0}) - \beta'_J \partial^J \delta \lambda_{I,0}\\
	&= -k_I \lambda_I -k_I \delta \lambda_{I,0}^{(1)} + \zeta\, \delta \lambda_{I,0}^{(1)} + \ord{\hbar^2}\\
	&= -k_I \lambda_I + 2 \delta \lambda_{I,0}^{(1)} + \ord{\hbar^2}\,,
	\end{split}
	\end{equation}
with $\zeta \equiv k_I \lambda_I \partial^I$. For the finite part of the \bef, it follows from the second term of~\eqref{eq:bare_coupling_invariance} that 
	\begin{equation} \label{eq:generic_2-loop_bef}
	\begin{split}
	\beta_I &= (\zeta - k_I) \delta \lambda_{I,1} + \delta \lambda_{J,0} \partial^J \delta \lambda_{I,1} +(\beta_J' \partial^J \delta \lambda_{K,0}) \partial^K \delta \lambda_{I,1} - \beta_J \partial^J \delta \lambda_{I,0} \\
	&= 2 \delta \lambda_{I,1}^{(1)} + 4 \delta \lambda_{I,1}^{(2)} + k_J \delta \lambda^{(1)}_{J,0} \partial^J \delta \lambda^{(1)}_{I,1} - (\zeta \delta \lambda^{(1)}_{J,0}) \partial^J \delta \lambda^{(1)}_{I,1} - 2 \delta \lambda_{J,1}^{(1)} \partial^J \delta \lambda^{(1)}_{I,0} + \ord{\hbar^3} \\
	&= 2 \delta \lambda_{I,1}^{(1)} + 4 \delta \lambda_{I,1}^{(2)} -2 \delta \lambda^{(1)}_{J,0} \partial^J \delta \lambda^{(1)}_{I,1} - 2 \delta \lambda_{J,1}^{(1)} \partial^J \delta \lambda^{(1)}_{I,0} + \ord{\hbar^3}\,,
	\end{split}
	\end{equation}
to two-loop order.\footnote{Note that $ (\zeta -k_I) $ picks up a factor of $ 2\ell $ when working on an $ \ell $-loop contribution with open index $ I $.}

\subsection{Running in theories compensated by evanescent counterterms}
We examine the running of physical couplings in the $ C $ scheme, where the evanescent operators are compensated by finite counterterms. This is a generalization of the argument of Ref.~\cite{Herrlich:1994kh} that evanescent couplings do not contribute to the running of the physical couplings in the $ C $ scheme. The running in the $ C $ scheme is compared to the $ S $ scheme of Section~\ref{sec:P-scheme_running}, wherein the evanescent couplings are simply eliminated and no finite counterterms are added. In a nutshell, we have 
	\begin{equation}
	S^{S}(g) \sim_\mathcal{P} S^{C}(g, \, \eta) = S^{S}(g) + \mathcal{E}_\mathcal{P} \,S^{C}(g, \, \eta) - \Delta^{\!S}\! S(g, \, \eta)\,, 
	\end{equation} 
where the two Lagrangians produce the same physics in the physical space. Here we wish to show that the running of the Lagrangian in the $ S $ scheme, cf.~\eqref{eq:2-loop_RG},
	\begin{equation}
	\beta_a^{S} = 2 \delta^S\! g_a^{(1)} + 4 \delta^S\! g_a^{(2)} -2 \delta^{S} \! \eta_{i,1}^{(1)} \partial^i \delta^{C}\! g_{a,0}^{(1)}\Big|_{\eta=0} + \ord{\hbar^3}\,,
	\end{equation}
is reproduced in the full $ d $-dimensional $ C $ scheme. Here and throughout, we take the superscript on the deltas to indicate what scheme the counterterm is taken in. At leading order, the physical counterterms are independent of the evanescent couplings, as insertions of evanescent operators increase the order in $ \epsilon $. Hence,
	\begin{equation}
	\beta^{C\,(1)}_a (g,\, \eta) = 2 \delta^{C}\! g_{a,1}^{(1)} (g,\, \eta)= 2 \delta^S\! g_{a,1}^{(1)} (g) = \beta^{S(1)}_a(g)\,.
	\end{equation} 
Before proceeding further, we introduce the shorthand notation 
	\begin{equation}
	\underline{f}(g,\, \eta) \equiv f(g,\, \eta) - f(g, \, 0)\,.
	\end{equation}
	
At the two-loop order, the equivalence of the physical \befs in the two schemes is not at all trivial. From~\eqref{eq:generic_2-loop_bef}, we have 
	\begin{equation} \label{eq:beta_R_2-loop_first}
	\beta^{C\,(2)}_a = 4 \delta^{C}\! g_{a,1}^{(2)} - 2 \delta^{C}\! g_{b,0}^{(1)} \partial^b  \delta^{C}\! g_{a,1}^{(1)} 
	- 2 \big(\delta^{C}\! g_{b,1}^{(1)} \partial^b  +
	\delta^{C}\! \eta_{i,1}^{(1)} \partial^i \big)  \delta^{C}\! g_{a,0}^{(1)}\,,
	\end{equation}
as $ \delta^{C}\! \eta_{i,0}^{(1)} =0  $ since no finite renormalization of the evanescent operators is introduced. It is a fundamental property of Feynman diagrams that the higher poles are related to the simple poles. This is incorporated in the finiteness of the  \bef.\footnote{Even if \befs can feature divergent terms in some cases, it is always possible to choose them finite~\cite{Herren:2021yur}.} In practical terms, the vanishing of the first pole in~\eqref{eq:bare_coupling_invariance} gives the generic condition 
	\begin{equation}
	2 \delta \lambda_{I,2}^{(2)} = \delta \lambda_{J,1}^{(1)} \partial^J \delta \lambda_{I,1}^{(1)}\,,
	\end{equation}
for the counterterms at two-loop order. This condition really speaks to the renormalized Feynman diagrams, so if we imagine a certain set of two-loop graphs, but now insert evanescent Lorentz structures in the vertices, the divergence will be reduced by one. Since the $ C $ scheme with its finite renormalization of the evanescent operators contain exactly the counterterms needed to cancel the finite physical part of any evanescent insertion, the pole relation will hold for the set of all graphs with evanescent operators but at reduced order of divergence. We have 
	\begin{equation}
	\begin{split}
	2 \underline{\delta^C\! g_{a,1}^{(2)} } 
	&= \underline{\delta^C\! g_{b,0}^{(1)} } \partial^b \delta^C\! g_{a,1}^{(1)} + 
	\delta^C\! g_{b,1}^{(1)} \partial^b \underline{\delta^C\! g_{a,0}^{(1)} } + 
	\underline{\delta^C\! \eta_{i,1}^{(1)} \partial^i \delta^C\! g_{a,0}^{(1)} } \\
	&= \delta^C\! g_{b,0}^{(1)} \partial^b \delta^C\! g_{a,1}^{(1)} + 
	\delta^C\! g_{b,1}^{(1)} \partial^b \delta^C\! g_{a,0}^{(1)} + 
	\delta^C\! \eta_{i,1}^{(1)} \partial^i \delta^C\! g_{a,0}^{(1)} - \delta^\mathrm{B}\! \eta_{i,1}^{(1)} \partial^i \delta^C\! g_{a,0}^{(1)} \Big|_{\eta=0}\,,
	\end{split}
	\end{equation}  
where we used that $ \underline{\delta^C \!g_{a,0}} = \delta^C \!g_{a,0} $, as these counterterms are introduced to compensate for the evanescent operators. Plugging this back into \bef formula~\eqref{eq:beta_R_2-loop_first}, it follows from $  \delta^{C}\! g_{a,1}^{(2)} =   \delta^S\! g_{a,1}^{(2)} +  \underline{\delta^{C}\! g_{a,1}^{(2)}}$ that 
	\begin{equation}
	\beta^{C\,(2)}_a = 4 \delta^S\! g_{a,1}^{(2)} - \delta^S\! \eta_{i,1}^{(1)} \partial^i \delta^C\! g_{a,0}^{(1)} \Big|_{\eta=0} = \beta^{S(2)}_a\,.
	\end{equation} 
Thus, not only do $ S^{C}  $ and $ S^S $ lead to the same action on the physical subspace, but the RG evolution of the physical couplings are equal. This is not a trivial result, considering that the two theories do not agree in $ d $ dimensions. Furthermore, this derivation demonstrates that once the theory has been cast in either the conventional $ C $ or  our $ S $ scheme, one can safely throw away the evanescent operators (and their counterterms), as they have no further bearing on physics or the physical RG. 

\section*{References}
{
\bibliography{References}
}
\end{document}